# A Scenario for Strong Gravity without Extra Dimensions

D. G. Coyne  (University of California at Santa Cruz)

A different reason for the apparent weakness of the gravitational interaction is advanced, and its consequences for Hawking evaporation of a Schwarzschild black hole are investigated. A simple analytical formulation predicts that evaporating black holes will undergo a type of phase transition resulting in variously long-lived objects of reasonable sizes, with normal thermodynamic properties and inherent duality characteristics. Speculations on the implications for particle physics and for some recently-advanced new paradigms are explored.

**Section I.  Motivation**

In the quest for grand unification of particle physics and gravitational interactions, the vast difference in the scale of the forces, gravity in particular, has long been a puzzle. In recent years, string theory developments [1] have suggested that the "extra" dimensions of that theory are responsible for the weakness of observed gravity, in a scenario where gravitons are unique in not being confined to a brane upon which the remaining force carriers are constrained to lie. While not a complete theory, such a scenario has interesting ramifications and even a prediction of sorts: if the extra dimensions are large enough, the Planck mass $M_P = \sqrt{\hbar c/G_N}$ will be reduced to $\sqrt{\hbar c/G_b}$, $G_b$ being a bulk gravitational constant $\gg G_N$. Black hole production at the CERN Large Hadron Collider is then predicted [2], together with the inability to probe high-energy particle physics at still higher energies and smaller distances. Experiments searching for consequences of the extra dimensions have not yet shown evidence for their existence, but have set upper limits on their characteristic sizes [3]. Experiments at LHC aimed at detecting such black holes are being planned [33], but suffer from not knowing exactly how a reduced-Planck mass black hole would behave.

Instead of a scenario, what is needed is a rigorous theory that can be unambiguously extrapolated to the low-energy scales of modern "high-energy" particle physics, and that can make detailed predictions for small black holes of masses near the pertinent Planck scale. While many physicists are currently laboring to find such a theory, it is a gigantic conceptual leap they are attempting to make. In the origins of modern quantum mechanics, before such a similar leap could be made, it proved useful to employ classical models with one or more simple *ad hoc* non-classical assumptions. This often gave insight as to whether such assumptions helped in solving a then-current dilemma and agreed with puzzling data existing at that time. The quintessential example of this is the development of the Bohr atomic model, where classical mechanics and electromagnetism, enhanced by the assumption of integer angular momenta, solved the problem of decaying electron orbits and correctly predicted the emission spectrum of simple atoms.

The analog for the quantum gravity problem would be to take a system which is adequately described by traditional particle physics and General Relativity (GR), but which breaks down at the Planck scale; similarly, one would try to patch up the "classical" treatment and resolve problems with minimal additional constraints. The project reported on below is such an attempt. The classical system analyzed is the evaporation of a black hole, and the new constraint is that the effective coupling of gravity changes radically as the Planck scale is approached. The principal result of this approach is that a completely different dynamical solution for the evaporation ensues. The reader may preview these results by examining Figs. 2-9, which show that the evaporation is eternal, free of physical infinities, possesses traditional thermodynamic properties after an apparent phase change, and likely conserves information. Duality under the exchange $M \to 1/M$ emerges naturally. This calculation does not quantize black hole states, so it fails in the full sense of the Bohr-like approach.

As a result of these calculations and a few further speculations, an alternative scenario emerges to explain the apparent weakness of gravity, without invoking extra dimensions. Again, it is a scenario without a complete theoretical structure, but has predictable consequences. In addition to those mentioned above, it implies that black hole production at CERN *certainly* will occur, in conventional 3+1 dimensions, but that the process will appear as ordinary business-as-usual particle physics.

To motivate this alternative scenario, consider some of the historical paths physicists have taken in the estimation of coupling constants. They were quite lucky with electrical forces, in that isolated electrons were readily available with which to measure the Coulomb force. In an extremely cold universe where neither atoms nor molecules could dissociate, the (non-human) physicists of such a world, until they discovered sub-eV accelerators, would have gauged the coupling strength of the electric force to be relatively weak and the radial dependence of the force going as $1/r^7$ – the van der Waals force. Only with deeper probing or with experiments on condensed matter would the unshielded Coulomb force have emerged. Our real world was similarly troublesome in the case of the strong force, where in fact its shielded nuclear form was interpreted as the complete coupling, and the strength and radial dependence of the true QCD coupling emerged later [4]. A *different* and *subtle* shielding mechanism was responsible for changing the real strength of the interaction to what has been termed "the nuclear van der Waals force", and for leading to asymptotically free particles at close range. Shielding mechanisms can mislead physicists, in both directions of the magnitudes of the couplings, until the mechanisms are fully understood. It is possible that with respect to gravity, the present situation is closely analogous to the imaginary "frozen universe" invoked above, wherein physicists have had no opportunity to properly sample the true strength of the gravitational coupling constant.

The fundamental assumption of the scenario of this paper is just that possibility: gravity is exceedingly strong, but it has never been observed at its true strength, not even in the strong limit of GR. It is hypothesized here that it exists in its unadulterated form, "hypergravity"[1], only within the interiors or on the horizons of small black holes, and that it is self-shielded by these complex new horizons. The scale of these structures might be of Planckian dimensions, and, as such, the nature of both the horizons and their interiors would be governed by quantum gravity. For this scenario, it is postulated that at this scale the horizon will not be the inescapable shroud of classical GR, but more like a leaky membrane, from which, at least locally, some information can emerge. At some number of Schwarzschild radii from this object, space-time would assume its classical GR form, as the "van der Waals" limit of hypergravity.

Such an hypothesis seems to fly in the face of previous observation or established physics, raising at least these questions:
1) Why is gravity seen at all from stars and planets, everyday objects and (presumably) from fundamental particles? Why would the basic underpinnings of cosmology (GR) depend on the leakage of gravity from small black holes, which, it is often suggested, are rare or may not even exist?
2) Configurations of electric or color charges can both have neutral states, but how can one neutralize (and thus shield) gravity?
3) Why should the existence of extra dimensions be doubted, when their natural appearance in string theory has led to the exact calculation of the expected entropy of black holes [21], as well as many other conceptual triumphs?
4) Finally, how could black holes be produced at CERN and not be noticed?

The end of Section III of this paper posits answers to these reasonable questions, based on calculations in Section II from a toy model illustrating how hypergravity could affect the final state of a small black hole and effectively shield itself. Section IV then allows further speculation based on the conclusions of Section III. Details of some calculations are given in appendices.

**Section II. Model**

*A. Classical portion*

As motivated above, one of the most puzzling outstanding questions in quantum gravity is how to extend the well-established theory of the evaporating black hole into the Planckian region, where the present model is expected to completely fail. In the treatment of this paper, it is first assumed that the *evaporation formula does not fail*. Then, to explore the idea of shielded strong gravity, a mathematical constraint is chosen to suggest how the gravitational coupling might appear in the limit as the black hole approaches Planckian proportions. A calculation combining the two descriptions can then be made. Only if the calculations predict that the physical variables stay within sensible bounds even where those of the classical process do not, can the approach be deemed self-consistent. This is the approach used; the candidate process spanning these environments is the Page-Hawking description of the evaporation of a Schwarzschild black hole [5]. A simplified version [6,7] adequate for these investigations will be used, as follows.

---

[1] Termed hypergravity, to distinguish it from the increasing effective strength of ordinary gravity in GR at small distances.

Hawking [8] showed that emission of particles from the vicinity of the black hole of mass M behaves as if the horizon were a thermal blackbody of temperature

$$T_{Bk} = \frac{\hbar c^3}{8\pi k_{Boltzman} G_N M},$$

albeit with corrections for the GR augmentation of the effective area of the horizon (twenty-seven times larger than the geometrical area) and suppression of long wavelength particles of non-zero spin because of the finite size of the radiator. $T_{Bk}$ is the temperature originally hypothesized by Bekenstein [9] on the basis of his extended second law of thermodynamics and the analogy between the equations of thermodynamics and black hole behavior [10]. The well-known equation of black hole evaporation [11] is then:

$$\frac{dM}{dt} = -\frac{\alpha(M)}{M^2}.$$

Considering the black hole as a radiating blackbody, the $1/M^2$ term comes from the product of the horizon area $4\pi(2G_N M/c^2)^2$ and the $T^4$ ($\sim 1/M^4$) factor of the ordinary Stefan-Boltzman law[2] for blackbody radiation. The $\alpha(M)$ lumps all the constants and gray-body corrections involved. It also reflects the increasing number of degrees of freedom available in the final state radiation as temperature $T$ increases, or, equivalently, as $M$ decreases. Inherent in this formulation is the assumption that at each temperature/energy scale of the hole, only particles fundamental at that scale are emitted. Fig. 1 from Halzen, et al. [7] shows this evolution of $\alpha(M)$ up through emission of the top quark and lowest Higgs states. It would continue to have steps at the threshold of production of each new particle. Thus, it will be constant if there is a "desert" free of fundamental particles up to $M_P$. It will double if a potential SUSY threshold is quickly passed in the evaporation process, and will decrease if there exists substructure to quarks involving fewer degrees of freedom [12].

---

[2] That law allows only two degrees of freedom for final state particles, e.g. the two polarization states of a photon; here, the α generalizes the law.

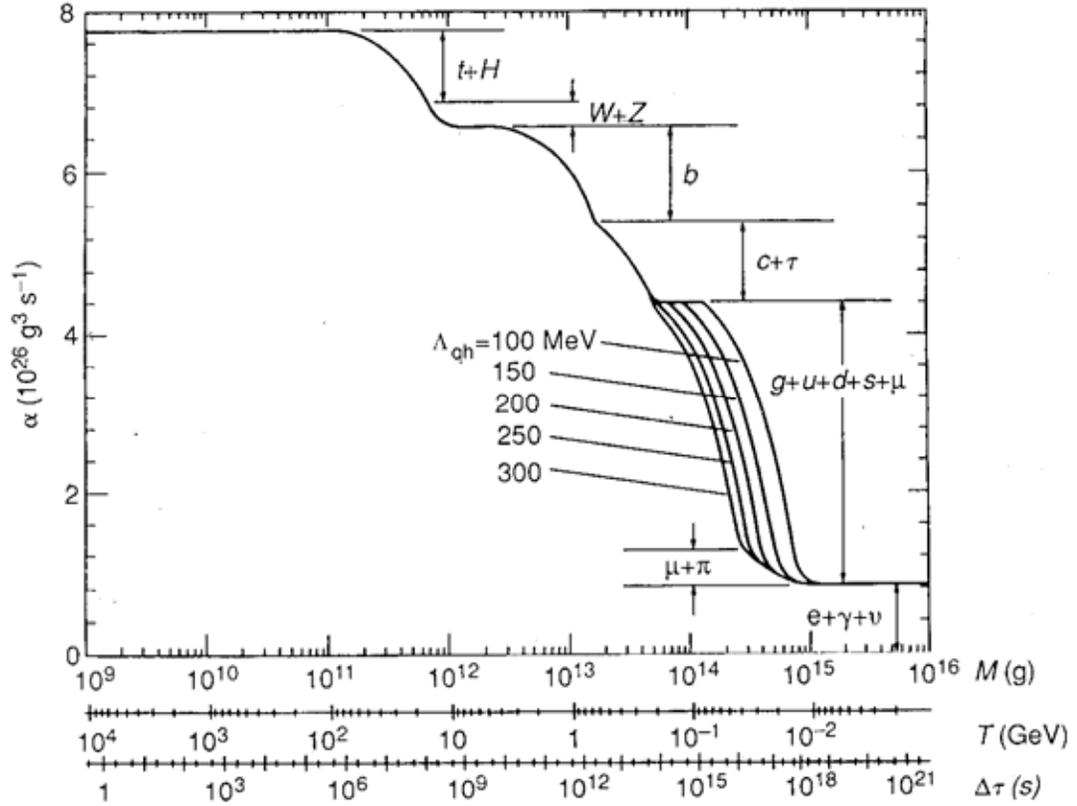

*Fig. 1 [Halzen, et al]: The parameter α counting the degrees of freedom of the Hawking mass radiation as a function of the black-hole temperature and lifetime. $\Lambda_{qh}$ is the quark-hadron deconfinement scale. The contribution from each particle species is indicated.*

In general, solution of the above equation is done numerically, though if one starts at $t=0$ with an initial black hole mass $M_o < 10^{11}$g and posits that no new particles exist (the desert hypothesis), then α is constant[3] and the differential equation is trivially solved to give:

$M(t) = (M_o^3 - 3\alpha t)^{1/3}$, which implies an absolute uniform evaporation time $\tau = M_o^3/3\alpha$;

Power output $= c^2 dM/dt = \alpha c^2 (M_o^3 - 3\alpha t)^{-2/3}$, implying a pole at $t = \tau$;

$T(t) \sim 1/M$, also implying an infinity at $t = \tau$;

$R_{(Schwarzschild)} = 2G_N M/c^2$, implying a vanishing horizon at $t = \tau$.

---

[3] Be aware that the more general case, with α free to vary, does not change the basic features of the evaporation, but simply changes the time scale on which those features are realized; this comment will be true for all models throughout this paper.

These Hawking time solutions will be shown in Fig. 2 to compare with the new model of this paper. Although these solutions are available as closed forms, the differential equation was also solved numerically to check the precision of non-analytic calculations to be done with the same program [17]. The agreement, for all practical purposes, was exact. The plots throughout this paper are shown with the graininess of the numerical integration to illustrate its resolution.

A feeling for the magnitudes involved using the constant (desert) value of α is provided by considering an initial mass of 1000 metric tons ($10^9$ g), for which the remaining lifetime is ≈ 0.5 s and the initial horizon radius is ≈$10^{-6}$ fm, radiating with instantaneous power ≈$10^{17}$ megawatts ($10^{30}$ erg/s) at an initial temperature $k_B T$ ≈10 GeV [13]. This is still in the valid portion of the formulation; the solution supposedly becomes invalid near the point at which the radius of the horizon of the black hole becomes of the order of the hole's Compton wavelength. This occurs at $T_P = T(M_P)$ ≈ 5·$10^{17}$GeV, at a time $M_P^3/3\alpha$ ≈ 4.6·$10^{-42}$s before final evaporation; this definition of $T_P$ is a factor of 8π smaller than the conventional one. Opinions vary about what happens at the remaining interval in which the formulation breaks down: whether or not remnants are found or if space-time reverts to being flat or has a horizonless singularity. For a historical vote (!) on this matter, see the discussion included with Ref. [14].

*B. A new constraint to the classical model and outline of the calculation*

This simplified model of black hole evaporation is used as the starting point for the model described below. The Page-Hawking equations are assumed to be valid everywhere, with the proviso that the strength of the gravitational coupling rises monotonically with increasing energy scale (temperature), eventually dominating all other known forces. This increase is not taken to be the usual logarithmic running of couplings [15], nor is it an expected scaling of *G* from dimensional arguments [16]. It is instead an *ad hoc* method of quantifying the approach of the evaporating hole to a region where it experiences the large and heretofore un-sampled large *G* hypothesized for the interior of small black holes. In order to guarantee that hypergravity will dominate over all other interactions, the following form is proposed for *G*:

$$G(T) = G_0 \frac{T_P - T_0}{T_P - T} \qquad (1)$$

Alternative forms are possible, or even preferable. Appendix C explores a more general and more tractable form for G which evolved later than **(1)**. The forms **(1)** and **(8)** are used in this paper because they show the essential common characteristics of a new mode of evaporation in a simple manner, free of arbitrary parameters.

The solution starts at $t_0 = 0$ at a mass $M_0$, where the temperature is $T_0$. By definition $G_0$ is the value $G$ has assumed at this initial $T_0$ and is constrained by $G_N$ — the value of $G$ at zero temperature (i.e., what is seen in our "frozen" universe):

$$G_N = G(0) = G_0 \frac{T_P - T_0}{T_P} \tag{2}$$

Equation **(1)** has a pole at $T = T_P$, on the assumption that this is the temperature where gravity absolutely dominates and where all particle physics merges into the black hole continuum. The solution continues by rewriting the evaporation equations to show explicitly their dependence on $G$, which for this model is no longer constant.

$$T = \frac{\kappa}{MG}, \tag{3}$$

where $\kappa = \hbar c^3 / 8\pi k_B$ = 8.17 $\cdot 10^{12}$ m$^3$s$^{-2}$K, or $k_B \kappa$ =1.06 $\cdot 10^{13} G_N$ (in units of GeV·g). Then

$$\frac{dM}{dt} = -\frac{\beta}{(MG)^2}, \tag{4}$$

where $\beta = \alpha G_N^2$ = 7.8 $\cdot 10^{26}$ g$^3$/s $\cdot G_N^2$, and

$R_{Sch} = 2GM/c^2$, the Schwarzschild radius, generalized to variable G. $\tag{5}$

$$M_P = \sqrt{\hbar c / G_N} \approx 22\mu g; \quad T_P = \kappa / M_P G_N \approx 4.8 \cdot 10^{17} \text{GeV} \approx 5.6 \cdot 10^{30} \text{ K} \tag{6}$$

With this formulation the evaporation equation **(4)** can be solved analytically, but is not quite so trivial since $G = G(T)$ and $T = T(MG)$. There is, however, an internal contradiction in this formulation, which must be corrected before proceeding.[4] $T_P$ was used as the expected energy scale at which the interior of a quantum black hole would experience overwhelmingly large gravity. But the lesson of Giddings, et al. [2] was that large $G$, *for whatever reason it occurs*, redefines the pertinent Planck scale. The Planck scale of that model was $M_P = \sqrt{\frac{\hbar c}{G_B}}$ and the one pertinent to this model is

$$M_{Pv} = \sqrt{\frac{\hbar c}{G}}, \text{ with Planck temperature defined as } T_{Pv} = \frac{\kappa}{M_{Pv} G}. \tag{7}$$

---

[4] This meandering presentation is meant to emphasize that it is *this* alteration which leads to an entirely new concept of "Planck Mass". The solution is totally uninteresting without this change.

The subscript $v$ serves to remind that the scale is now a variable one, depending on the value to which $G$ converges in the solution (if indeed it does at all). Thus $G(T)$ is redefined in **(1)** as:

$$G(T) = G_0 \left[ \frac{T_{Pv} - T_0}{T_{Pv} - T} \right] \tag{8}$$

(This particular model will be called the "variable pole model", or used interchangeably with the term G-pole or even "hypergravity" — though presumably other models of strongly-shielded gravity could exist).

Does this redefinition **(8)** mean that the traditional Planck scales $M_P$ and $T_P$ play no role? A slight manipulation of the above equations shows that they could play an important role. Using the definitions **(7)** inserted into equation **(8)**, one finds:

$$G(T) = G_0 \left[ \frac{T_P - T_0 \sqrt{\frac{G}{G_N}}}{T_P - T \sqrt{\frac{G}{G_N}}} \right] \tag{9}$$

The constraint on $G_0$ found in **(2)** still holds identically, since $G(0)/G_N = 1$. Thus the implicit equation defining $G$ still references the Planck scale as special, but says that the temperature at which $G$ appears to become infinite (the pole position) is at $T_{P_v} = T_P \sqrt{G_N/G}$, or equivalently, at $M = M_{P_v} = M_P \sqrt{G_N/G}$. The reason why $T_P$ is still important will become clear only with the details of the solution of equation **(4)**, which will *dynamically* determine $G$ and the interesting scale of black hole evaporation. This solution can be found by a series of analytic steps and numerical calculations, outlined below.

Equation **(3)** can be inserted in equation **(9)** to give a quadratic in variables $\sqrt{G}$ and $M$, plus constants. Identities useful for simplification are

$$T_P/(T_P - T_0) = G_0/G_N, \; T_P/T_0 = (M_0 G_0)/(M_P G_N) \text{ and most}$$
$$\text{obscurely, } G_0/G_N = 1 + M_P/M_0 . \tag{10}$$

The solution for the quadratic is

$$2 M_0 \sqrt{G} = M_P \sqrt{G_N} \left\{ \frac{M_0}{M} - 1 \pm \sqrt{\left(\frac{M_0}{M} - 1\right)^2 + 4 \frac{M_0}{M_P}\left(1 + \frac{M_0}{M_P}\right)} \right\} . \tag{11}$$

The ambiguity in sign seems spurious, because in the limit $t=0$, $M=M_0$ and if we take $M_0 >> M_P$, then $\sqrt{G} = \pm\sqrt{G_N}$. No negative sign is needed to connect to the expected Hawking solution, which in this region demands that $G = G_N$. The first order differential equation **(4)** can now be solved entirely in terms of the one independent variable M by substituting the expression **(11)**. The time integral is explicit and can be numerically evaluated:

$$\frac{\beta t}{G_N^2} = -\frac{1}{16}\left(\frac{M_P}{M_0}\right)^4 \int M^2 \left\{\frac{M_0}{M} - 1 + \sqrt{\left(\frac{M_0}{M} - 1\right)^2 + 4\frac{M_0}{M_P}\left(1 + \frac{M_0}{M_P}\right)}\right\}^4 dM \quad (12)$$

Alternatively, the differential equation **(4)** can be solved numerically [17]. Then equations **(11)**, **(3)**, **(5)** and **(4)** can be used to calculate $G(t)$, $T(t)$, $R(t)$ and power output, respectively. Note that for M and $M_0$ both $>> M_P$ and $M = O(M_0)$, the integral reduces to $\beta t/G_N^2 = -\int M^2 dM$, exactly the Hawking form.

*C. Details of the solution for the variable pole model*

A few general features of the solution are evident by inspection. Since $\beta$, M and G are all positive definite, $M(t)$ can only decrease with time. As noted above, a typical solution starting at large initial mass begins by mimicking the standard Hawking behavior, as if the object had a finite life. But for very small M, one sees that $G \to G_N M_P^2/M^2$ and thus, using **(4)**, $M \to \propto 1/t$: the black hole will never evaporate and can reach arbitrarily small masses. $G(t)$ grows quadratically with time, never reaching a pole: it is dynamically suppressing its own infinities. Temperature $T \to \propto 1/t$ as well: the black hole cools off! (The proof of these and other asymptotic behaviors that demonstrate the inherent duality of the model under the exchange $M \leftrightarrow 1/M$ can be found in Appendix A). Clearly, something interesting has happened in the intermediate region where Hawking behavior has been transcended.

From a numerical solution to the differential equations, the following Figs. 2a-d illustrate these initial conclusions and introduce new surprises. These figures do not display the vast majority of the time scale (for $M_0$ starting $>> M_P$), because the solution in that time region is indistinguishable from the Hawking case. Instead, the figure begins at $M_0 = 3 M_P$, where a sudden change from the traditional behavior soon ensues. Note that the time intervals of pertinence are very small, of $O(10^{-40}s)$, but are still $>> \tau_P = 10^{-44}$ s by four orders of magnitude. At $t_c = t - t_0 \approx 2.7 \cdot 10^{-40}s$, $M \approx 1.33 M_P$, the black hole temperature (Fig. 2a), which has slowed in its increase toward a Hawking-like pole, abruptly peaks and then drops precipitously, as does the power output of the hole.

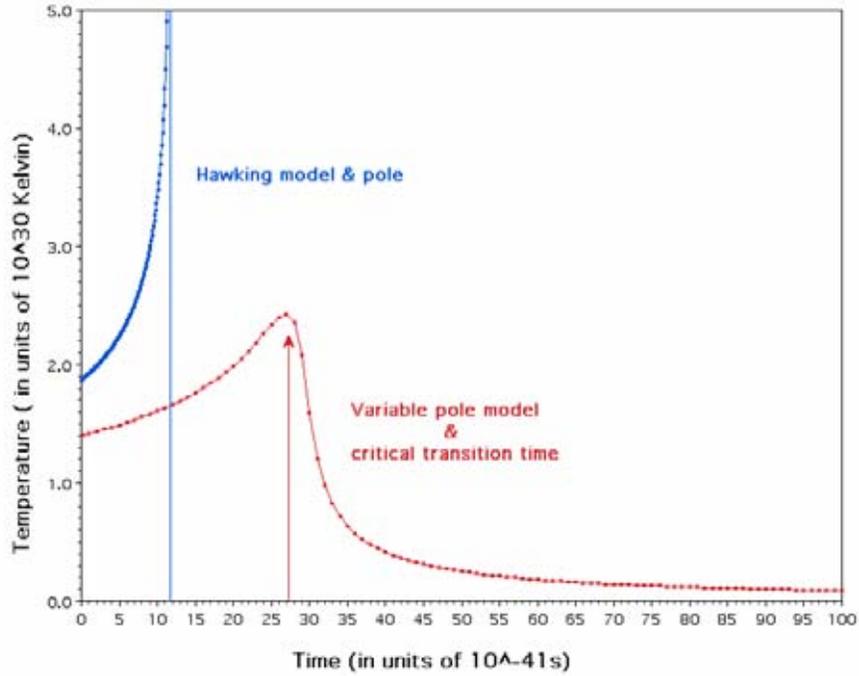

Fig. 2a: *Comparison of temperatures in black hole evaporation.*

The mass (Fig. 2b) continues its usual plunge toward zero, but then only about $0.5 \cdot 10^{-40}$ s after $t_c$, at a mass of about 0.1 $M_P$, starts to change to the $1/t$ behavior. For all practical purposes, it reaches asymptopia soon thereafter.

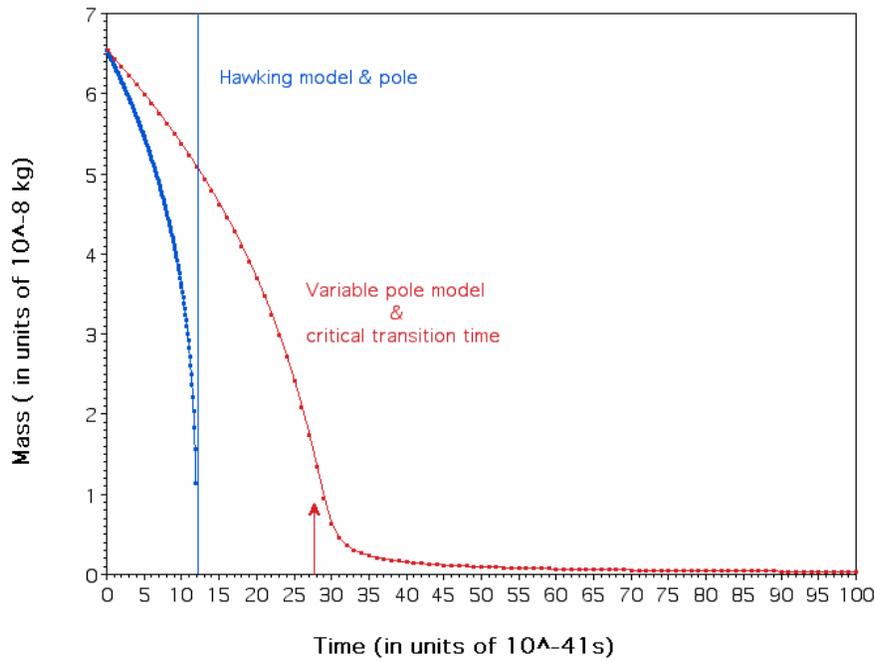

Fig. 2b: *Comparison of mass in black hole evaporation.*

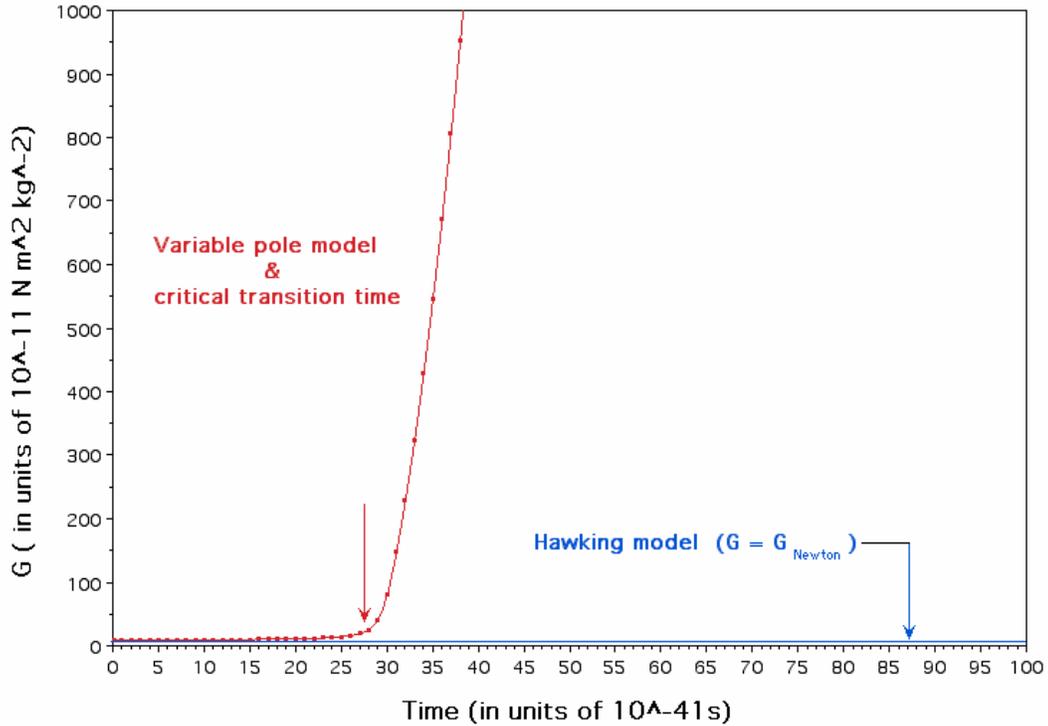

Fig. 2c: Comparison of gravity strength in black hole evaporation.

The value of $G$ (Fig. 2c), hovering at a few times $G_N$ up to the critical transition, starts to climb. At this transition point it is roughly $5G_N$, appreciably stronger, and is beginning to show the pole-like behavior built into the model. It would be natural to expect an infinity in the value of G as M crosses the value $M_{Pv}$. Closer examination (Table 1 and Appendix A) shows that the pole $M_{Pv}$ starts to run away to lower values from whatever mass the black hole has achieved, but is caught at $M = M_P/4$. At exactly this point, the "pole" is suppressed by a zero of the numerator of equation **8**. Thereafter, the "pole" is locked to a value just larger than $M$ by $\delta$, and $\delta \to 0$ as $t \to \infty$.

An unanticipated surprise is the behavior of the black hole horizon (Fig.2d), which of course in the old Hawking solution would quickly go to zero radius, introducing conceptual problems: an object being incredibly smaller than its Compton wavelength; a naked singularity; possibly lost information. Here, the solution to this model halts the shrinkage of the horizon at $R = 0.8 \cdot 10^{-32}$ cm, five times the Planck length, and then proceeds to re-inflate the horizon! Since $R$ is proportional to $GM$, its asymptotic form in this time region is $\propto t^2 \cdot (1/t) \propto t$.

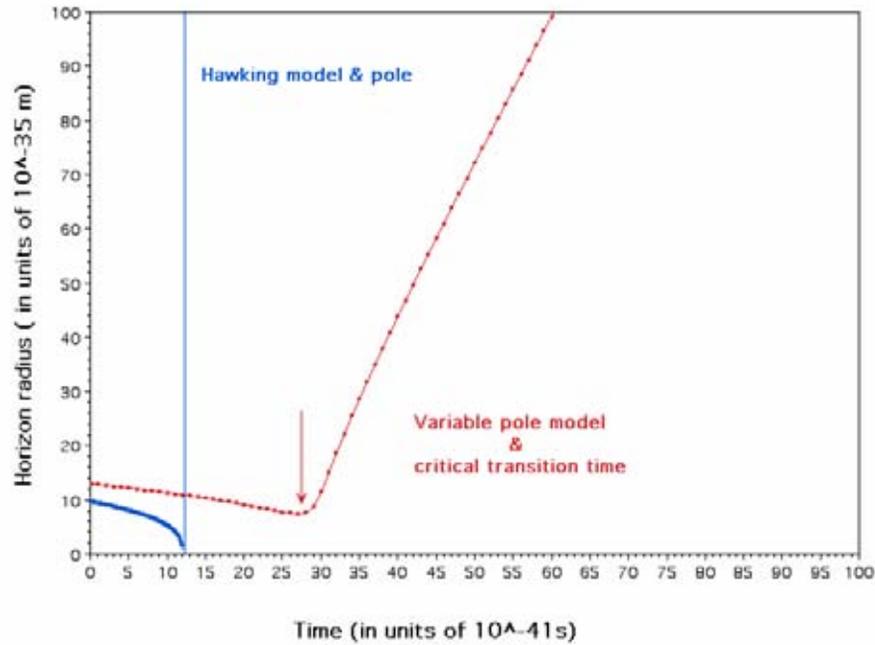

*Fig. 2d: Comparison of horizon size in black hole evaporation.*

Fig. 3, below, shows all of the variables of the new model on a single plot, where each variable is shown relative to the Planck units of that quantity. This display is useful to see the correlated behavior of the variables. (In this figure, the "Planck power" referred to is the power output at the Planck mass of the ordinary Hawking model. It is enormously suppressed in the new model). The inflation of the horizon is seen to be time-correlated with the sudden transitions in behavior of the other variables.

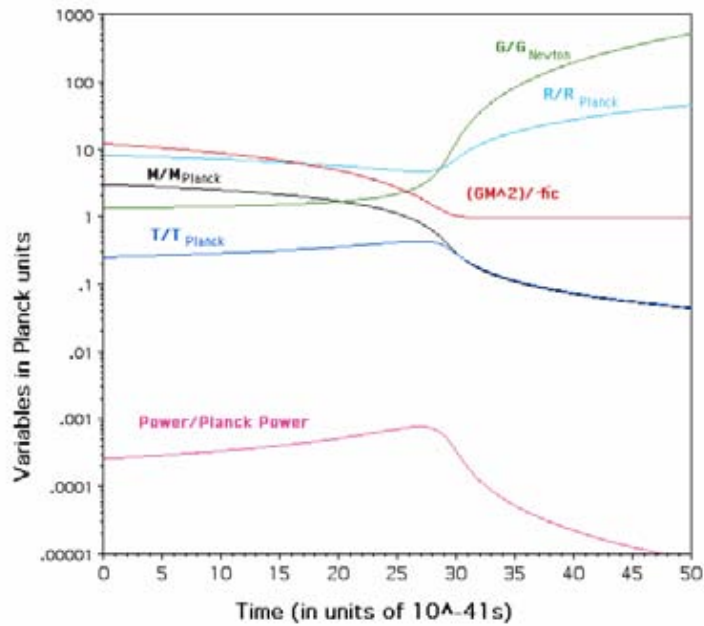

*Fig. 3: Behavior of the variables relative to their Planck values*

What is the significance of this linear inflation of the "new horizon"? Figs. 3 and 4 show that at the transition time the quantity $GM^2$ becomes essentially constant at the value $\hbar c$. This is exactly what is needed to keep the horizon radius $2GM/c^2$ greater than the Compton wavelength $\lambdabar = \hbar/Mc$ of the remaining black hole:

$$2GM/c^2 > \hbar/Mc \Rightarrow GM^2 > \hbar c/2$$

Thus, after the transition, the physical size of the object never violates a reasonable limit, quite unlike the constant $G_N$ Hawking solution.

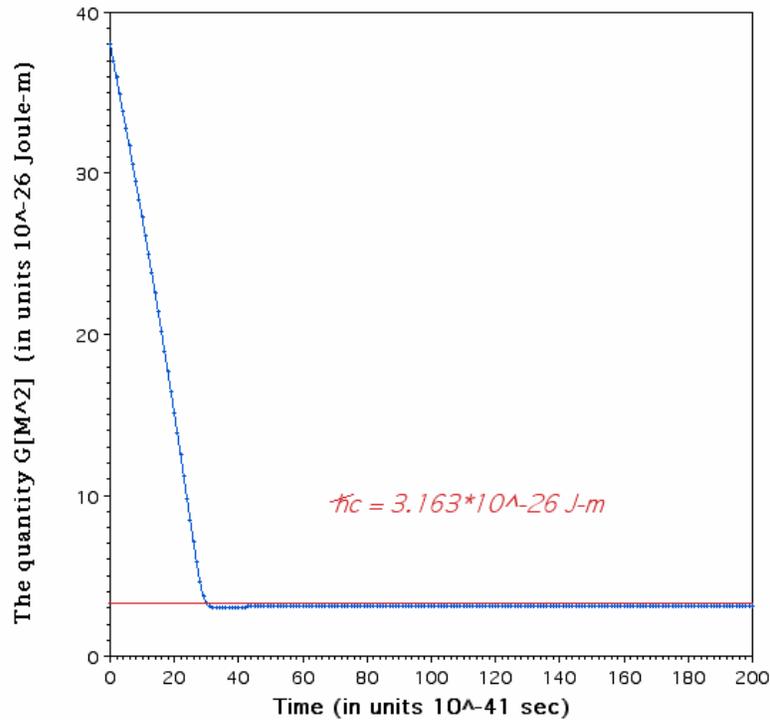

Fig. 4: Transition of the black hole to an object of Compton wavelength proportions. The quantity $GM^2$ vs. time for the variable pole model is shown.

There is even more significance to the inflation of the horizon. Figs. 5(a-f) show the time evolution of the embedding diagram of this system relative to that of a traditional Hawking solution. (Hawking solution on left, variable-pole on right; the radial direction is in the x-y horizontal plane, with the diagram terminating at the horizon. The embedding "funnel" is also cut off at an arbitrary positive constant of z and r to better illustrate its shape). At a mass 10 $M_p$, the solutions are indistinguishable. The horizon of the variable-pole solution shrinks more slowly than does that of the Hawking solution, as $M \to M_p$, and its corresponding potential well at a given radius is deeper. (The embedded surface is *not* the effective two-dimensional potential, but is intimately related to it through the appearance of the lapse function both in the metric and the potential [22]. Plots of the potential look very similar, so would be redundant here.)

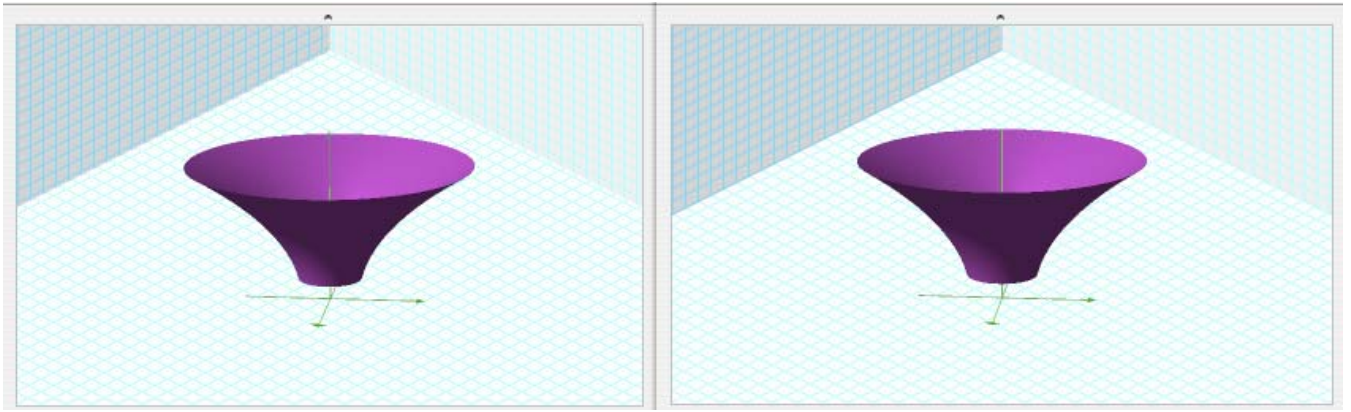

*Fig. 5a: Hawking and variable pole black holes of mass 10 $M_P$*

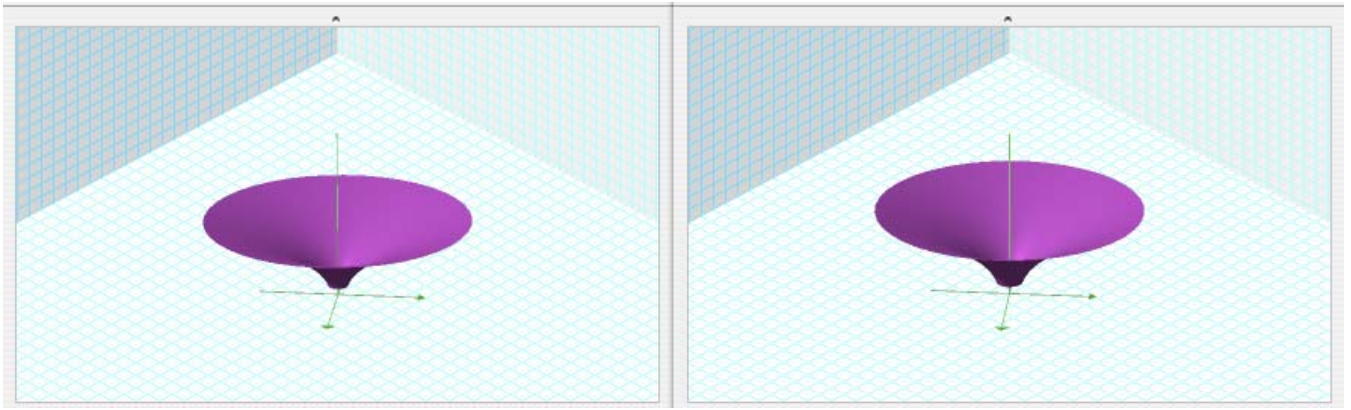

*Fig. 5b: Hawking and variable pole black holes of mass 3 $M_P$*

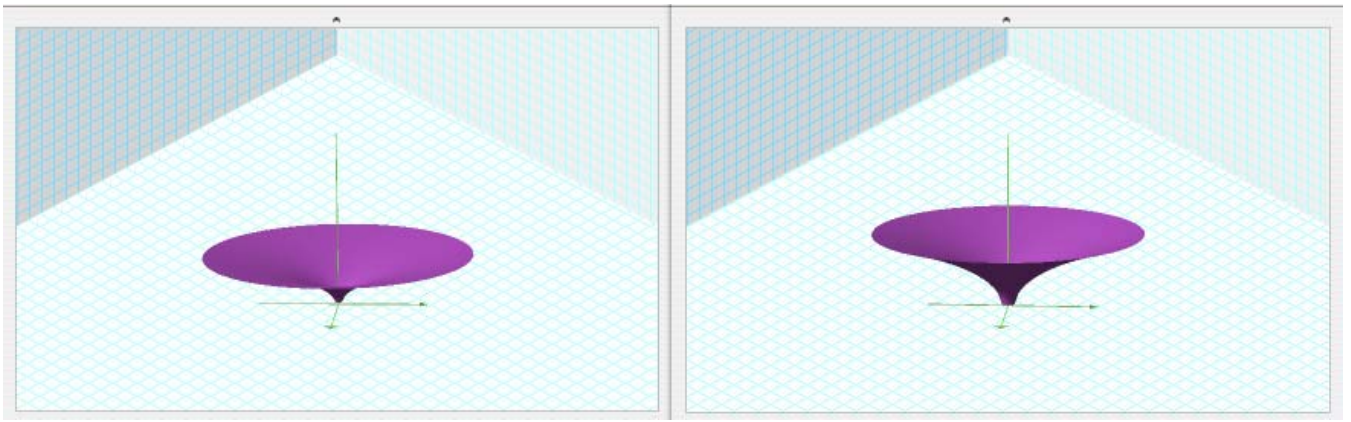

*Fig. 5c: Hawking and variable pole black holes of mass 1.1 $M_P$*

For $M < M_P$, the Hawking solution has an ever-shrinking horizon and a shallower but increasingly sharper potential well, characteristic of its problems of zero gravitational potential surrounding a naked singularity. The variable pole solution reverses these trends and inflates the horizon, deepening and broadening the potential well.

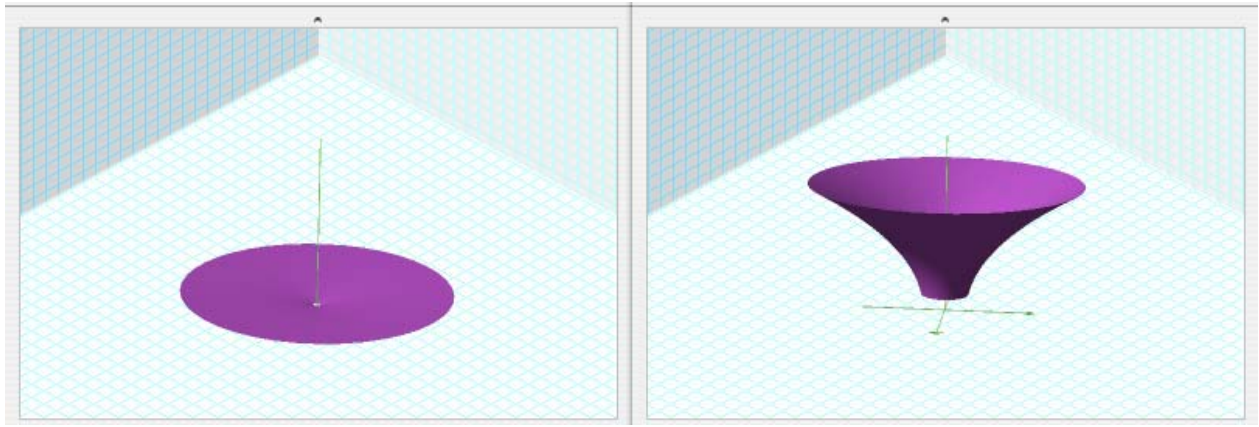

*Fig. 5d: Hawking and variable pole black holes of mass 0.122 $M_P$*

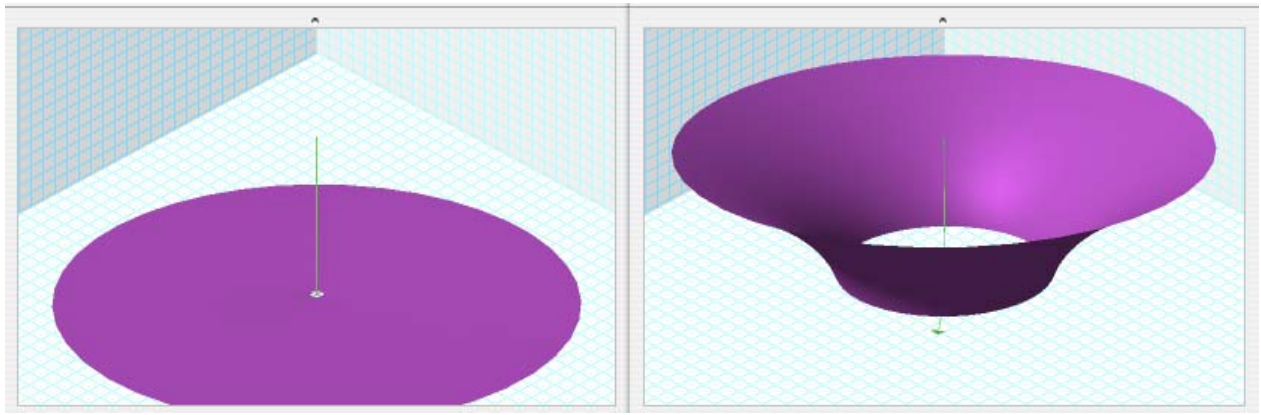

*Fig. 5e: Hawking and variable pole black holes of mass 0.025 $M_P$*

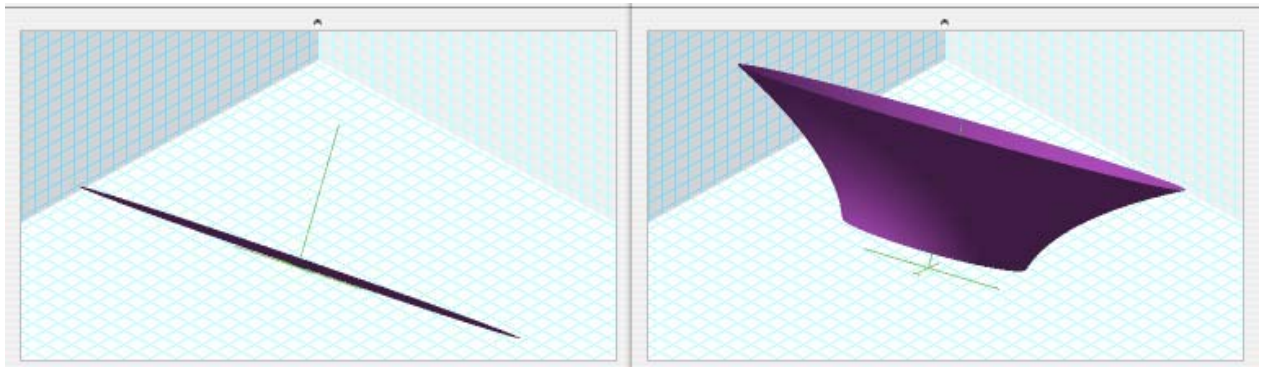

*Fig. 5f: As in Fig. 5e, but tilted to show profile*

The variable *G*-pole model inflates the horizon and keeps the singularity "clothed", within a deep potential well. Depending on the nature of the horizon and its interior, quantized stationary states are at least plausible, as will become more evident as the model is investigated. It is these two key features that will help make the model consistent with the idea that a shielded large gravitational force could lead to particle-like final states.

A first step on the way to see this, ironically, is to recast a graph due to Carr and Giddings [18] that shows, traditionally, why elementary particles can't be black holes. Fig. 6a shows the plot of size vs. mass with the regions forbidden by classical GR and by QM. As the mass falls below the Planck mass, the GR limit is superceded by the rising QM limit, and one concludes that particles must be much larger than their GR black hole size. In the sub-Planckian region, a mass can't be compressed to its black hole size because its density is limited by its Compton wavelength; if it were squeezed that tightly, the uncertainty principle says that its mass-energy would increase and it would translate far to the right. Carr and Giddings then describe how larger black holes might be produced in laboratory collisions if extra dimensions of sufficient size do indeed exist.

The variable-pole scenario, without extra dimensions, negates this traditional conclusion and also leads to a different view of laboratory production than that promoted by these authors. Fig. 6a also shows that the radius in the variable-pole solution for masses < $M_P$ easily evades the QM limit, tracking it exactly. That allows, but not compels, elementary particles to be semi-stabilized black holes. Note that Fig.6a also demonstrates the duality property of the horizon size under the exchange $M \leftrightarrow 1/M$.

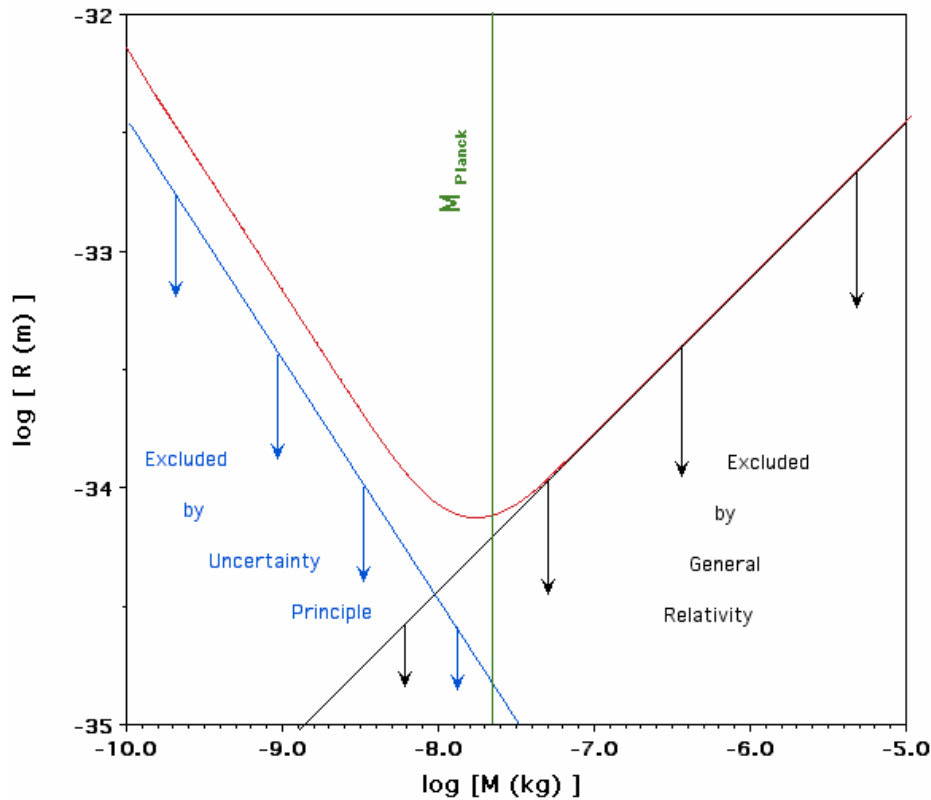

Fig. 6a: In the vicinity of the Planck mass, the excluded regions for physically reasonable particles: black line (positive slope) is Hawking hole and GR limit; blue line (negative slope) is UP limit; red curve is a variable-pole hole.

Following the same authors, the mass density required to produce a black hole is calculated for both classical GR and for the variable pole model, and is displayed below on Fig. 6b. One sees that the classical GR density goes like $1/M^2$ and surpasses the authors' "natural" limit of $10^{97}$ kg/m$^3$ just below $M_P$. The density required of this particular hypergravity model goes like $M^4$ below $M_P$ and it becomes ever-easier, as $M$ decreases, to form holes, never violating that limit. But these holes produced could then be just those scalar elementary particles that populate the discrete spectrum in the region of available center-of-mass energy. As with traditional particle physics, one can form an intermediate virtual state with arbitrary mass, but it materializes in a collection of ordinary quantized states. Details of the evaporation signature will be discussed later in Section III.

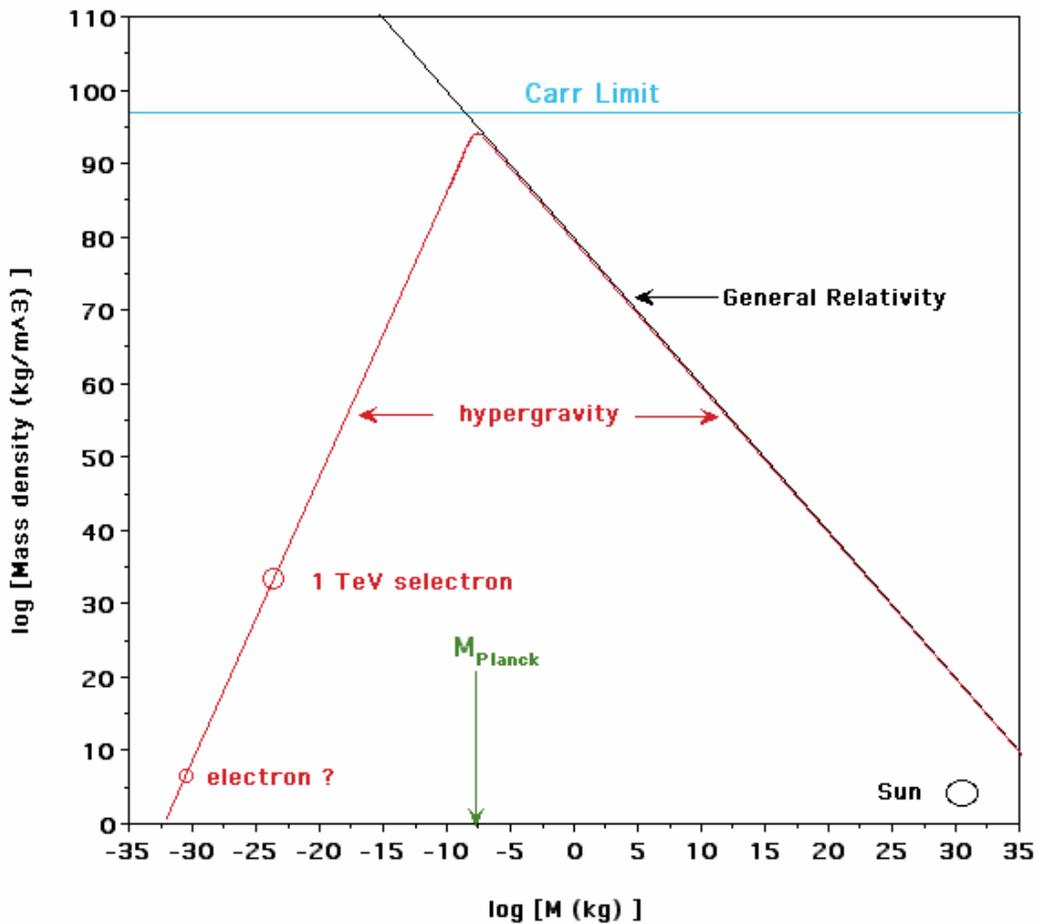

Fig. 6b: Mass density required to produce a black hole; GR/Hawking is negatively sloped black line, while hypergravity is both lines connected by the red curve.

## D. Thermodynamics of the variable pole model: specific heats

Additional insights to this solution can be gained by examining the behavior of the evaporation as a non-equilibrium thermodynamic system. Black hole thermodynamics is typically described [19] by intensive variables $T$ and angular velocity $\Omega$, and by extensive variables entropy $S$ and angular momentum $J$. The First Law provides a constraint:

$$dU = d(Mc^2) = TdS + \Omega dJ.$$

For a Schwarzschild black hole, the $\Omega dJ$ term (analogous to $PdV$ classically) vanishes, and it is sensible to discuss only the quantities $U$, $T$, $S$ and the specific heat $c_{BH} = \dfrac{dU}{UdT}$. The thermodynamic properties revealed will be grossly different than for a traditional Hawking evaporation, as has already been demonstrated for the temperature.

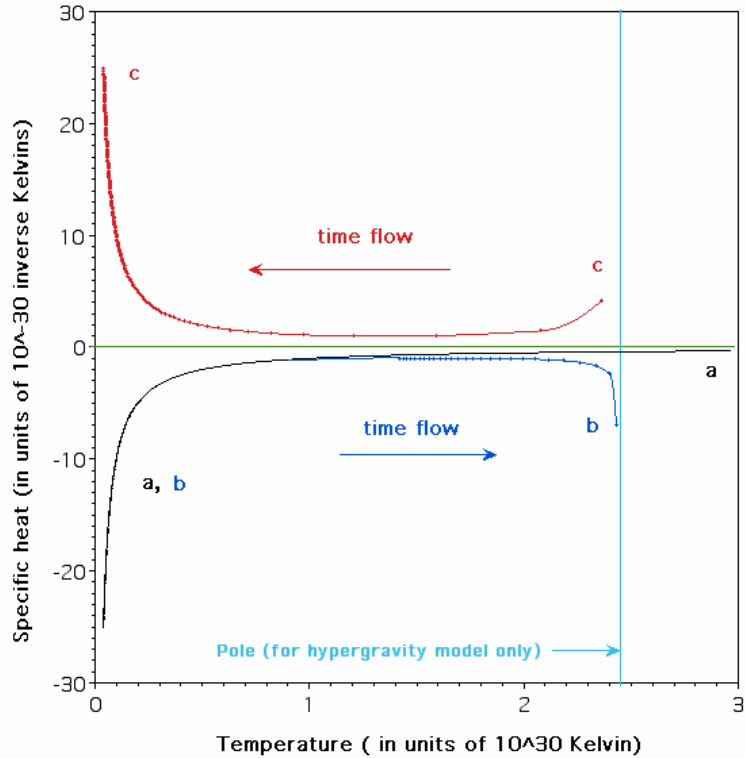

Fig. 7: Comparison of specific heats in hole evaporation: a) black lower branch is Hawking hole; b) blue/black lower branch is hypergravity hole; c) red upper branch, for hypergravity only.

The specific heat of a classical black hole is always negative, as shown in Figs. 7 and 8 (arrows indicate the direction of time evolution in the evaporation process). However, this new type of black hole evaporation has a surprise: the specific heat has a pole in $T$ (or in $U$) at the "transition" temperature, and, after the infinite discontinuity, a second branch in which its specific heat is normal and positive (and where the black hole can be in equilibrium with a thermal bath). It is tempting to interpret this discontinuity as a phase transition, but from what, into what? There have been conjectures that black hole horizons really have substance, as found in theories in which they are emergent phase transitions of the vacuum, or simply shells of emitted particles [20]. Here, all that can be deduced is that there is a critical point where the hypergravity model ceases to follow the unusual thermodynamics of the classical black hole, and instead produces a state that looks more and more like an object obeying traditional thermal physics.

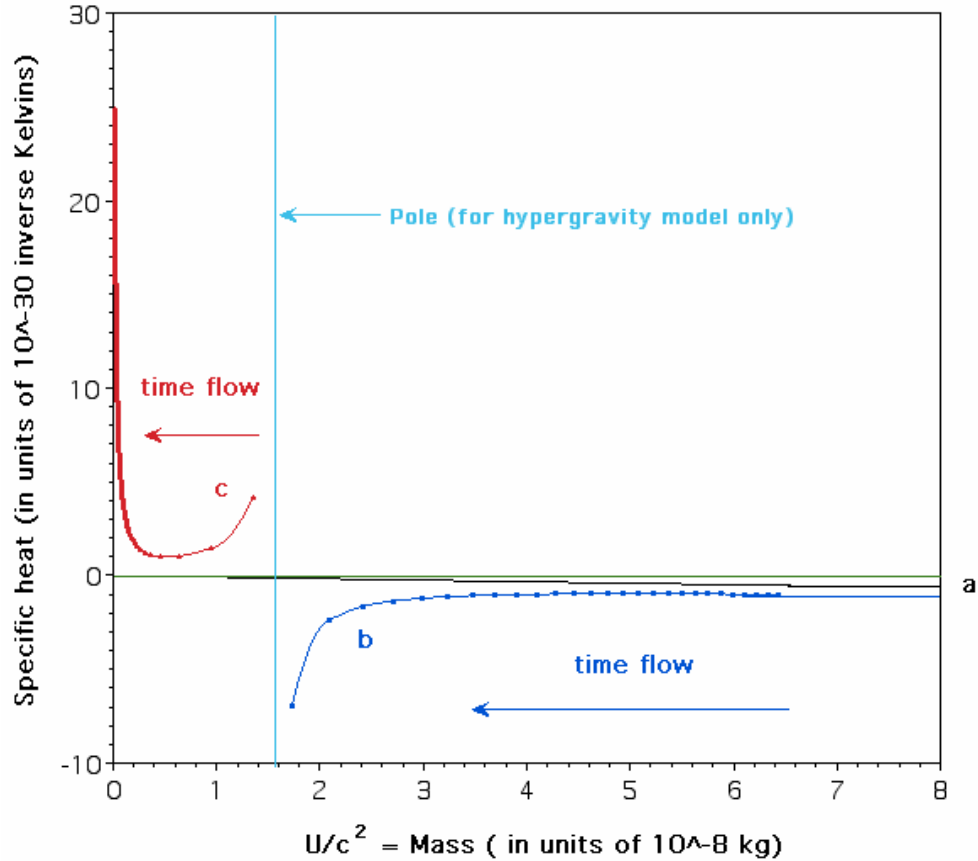

Fig.8: As for Fig. 7 except plotted vs. mass

*E. More thermodynamics: entropy*

In any discussion of black hole thermodynamics, entropy is considered to be one of the most important properties, given its close connection to information flow and the quantum mechanical concept of pure states. From thermodynamic analogies, it is expected to be: **S** $\equiv S/k_B = c^3 A/(4G\hbar)$ [$A$ = horizon area], and thus equal to $4\pi M^2 G/(\hbar c)$, taken as a measure of the concealed information within or on the horizon. The entropy of black holes is without experimental confirmation, but is so well-established theoretically that the agreement of the string theory calculation [21] with the standard formula has been taken to be a major triumph for that theory. Numerical evaluation of the formula reveals that entropy can be huge for astrophysical black holes: a solar-mass black hole has **S** = 1.05·10$^{77}$, a factor of 10$^{12}$ greater than the entropy of the Sun itself [22]. By contrast, the same formula, applied much lower than its expected region of validity, gives the entropy of a Hawking black hole of Planck mass to be 4π, tending to zero like $(M/M_P)^2$. The variable-pole model, applied to this sub-Planckian region, predicts that $M^2G$ becomes constant at $\hbar c$, which (coupled with the classic entropy formula) might be taken to suggest that these objects have universal asymptotic entropy 4π, or information content **S**$_{Shannon}$ = **S**/ln2 ≈ 18 bits[5].

However, both the traditional formula for entropy and its applications as above are invalid for the variable-pole model. The ordinary expression for entropy of a black hole is derived from the relations

$$c^2 dM = dU = T_{Bk} dS = \frac{\hbar c^3}{8\pi GM} \frac{dS}{k_B} \equiv \frac{\hbar c^3}{8\pi GM} d\mathbf{S},$$

where the units of $S$ are joules/Kelvin and those of **S** are nats, or (sometimes) nits; **S**/ln2 gives bits. For constant $G = G_N$, integration over d**S** of this expression gives the usual **S** = $4\pi M^2 G/(\hbar c)$. A usually unstated assumption is that **S**$(M=0)=0$, or alternatively, that for sub- or near-Planckian objects, none of the classical approach means anything.

For the variable-pole model, care must be taken to avoid unwarranted assumptions about what entropy might be from this perspective. For the region with $M \gg M_P$, the integration $\Delta\mathbf{S} = \int_{M \gg M_P}^{M \gg\gg M_P} d\mathbf{S}$ gives the classical solution, since in this region $G(M) = G_N$ to any arbitrary precision. The other asymptotic region of interest is for $M \ll M_P$, where the variable pole model does not fear to tread. There, with $M^2G = \hbar c$ to arbitrarily good precision,

$$\Delta\mathbf{S} = \int_{M_L \ll\ll M_P}^{M_U \ll M_P} \frac{8\pi M}{\hbar c} \cdot \frac{\hbar c}{M^2} dM = 8\pi \ln \frac{M_U}{M_L}$$

This means that a logarithmically increasing entropy is expected in this region. A constant of integration has to be added to define the absolute entropy at some small value of $M = M_L$. Since no scalar particle has ever been found up to a conservative limit of $M = 100$ GeV [24], $M_L = 100$ GeV (and $S_L = 0$) are taken to define the lower limit of the entropy. (If it is desired to have some constant entropy $S_L \neq 0$ for this lowest scalar, that will just be an additive constant for the graphs shown below).

---

[5] Curiously, this is almost the same information content (17 bits) required to reside in each single atom of a monatomic ideal gas, described by Maxwell-Boltzmann statistics, in order to account for the thermal entropy of the gas [23].

The global solution for the absolute intrinsic entropy of a black hole/particle is then calculated using a potpourri of the two asymptotic solutions and a numerical integration in the intermediate transition region. It is shown in Fig. 9a for both the classic GR case and the $M_L = 100$ GeV, $S_L = 0$ version of the variable pole model. In addition, for the purpose of seeing the effect of the boundary condition, a case where $M_L = 10$ eV (!) and $S_L = 0$ is shown. Note that $\Delta S$ taken between any two masses is independent of the additive constant, which basically only defines the endpoint of the evaporative process and remains a parameter of the model.

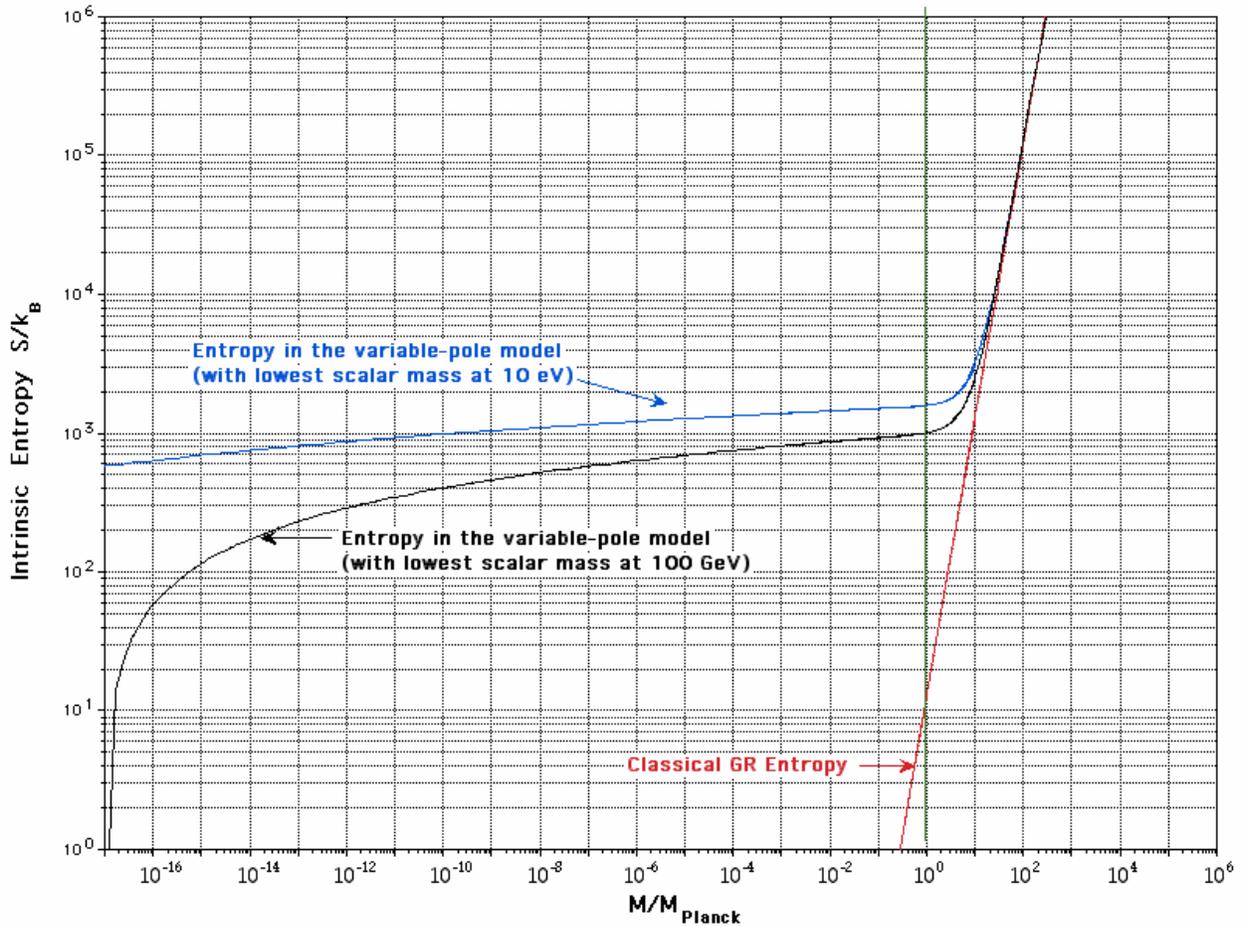

Fig 9a: Comparison of entropies of black holes/particles for varying models

Thus, the model predicts that sub-Planckian objects have appreciable entropy, and much more so than extensions of the classical model. Even at $M = M_p$, it exceeds the (dubious) classical value by two orders of magnitude. This might be thought to be unimportant in the super-Planckian physics of black holes, where the additive constant (about 1000) is completely trivial. But the view espoused here—that the sub-Planckian objects are particles—leads to the conclusion that large black holes with entropy decay into many smaller ones, which also have intrinsic entropy. It turns out that the logarithmic behavior found for entropy in the sub-Planckian region, coupled with the usual behavior above the Planck mass, effectively *guarantees* the non-violation of the second law of thermodynamics, *just* for the intrinsic entropy of black holes (An example is derived in Appendix B). This in turn must bear on the infamous "information problem", even though many believe that has been partially or completely solved by other considerations.

In discussions of black hole entropy and quantum black holes, attention is often centered on the entropy per unit area as a critical parameter [25], related to observance of the Bekenstein Bound [26]. For the classic GR black hole, $\mathbf{S}/A = c^3/(4G_N\hbar)$ is constant, a value also found in some versions of string theory; QFT predicts infinite density, considered a disaster for that theory [27]. The variable-pole model gets the GR result for $M \gg M_P$, but for $M \ll M_P$ it is:

$$\frac{\mathbf{S}}{A} = \frac{8\pi \ln \frac{M}{M_L}}{4\pi \left(\frac{2MG}{c^2}\right)^2} = \frac{c^3}{2\hbar G} \ln \frac{M}{M_L} = \frac{M^2 c^2}{2\hbar^2} \ln \frac{M}{M_L}$$

Fig. 9b shows the global solution for entropy density with $M_L$ = 100 GeV, $\mathbf{S}_L$ = 0, illustrating these two limiting behaviors. There is a small bump in entropy density between the two asymptotic regions and near the Planck mass, which is of unknown significance; it briefly (and modestly) over-saturates the Bekenstein Bound.

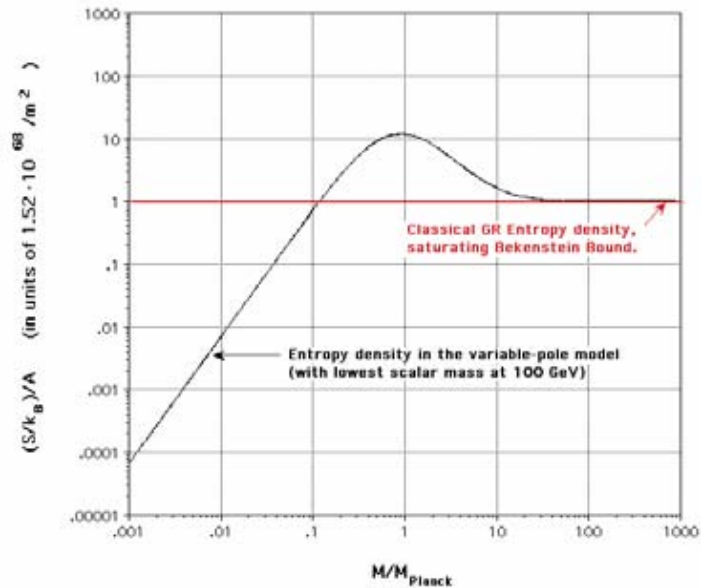

Fig 9b: Comparison of entropy densities of black holes/particles for two models

However, as $M \to 0$ and $G \to G_N M_P^2/M^2$ and the horizon inflates, it appears that black holes as elementary particles have very sparse information content per horizon area (becoming arbitrarily close to zero), but appreciable absolute information content. For example, a scalar black hole stabilized at $M$ = 1 TeV would have $\mathbf{S}_{Shannon} \approx$ 87 bits, allowing (but not requiring) about $10^{26}$ possible internal configurations of the microstates of its shielding horizon, all within a radius of about $2 \cdot 10^{-4}$ f. Because of the logarithmic dependence, all other particles, from electron-like masses almost up to the Planck scale, would have similarly large numbers of microstates; this would seem to provide adequate phase space for scalar fundamental particles with widely differing properties and enormously complex horizons. (At the Planck mass, the 1440 bits provide $10^{434}$ possible configurations!) Other ramifications of this information content will be explored in Sections III, IV and in Appendix B.

**Section III. Discussion**

At first glance the scenario derived in the last section seems bizarre, but it is not: this is exactly what would be expected if an evaporating black hole leaves a remnant consistent with quantum mechanics. One might posit that the black hole smoothly turns into something approximating a large and unstable elementary particle, which then continues to evaporate (decay) into familiar stationary states. Many have speculatively done so: some models suffered from unrealistically small horizon sizes, and some have found ingenious ways to avoid them [28]. No model has provided reliable predictions of particle spectra. One early bold speculation put the matter quite simply:

> At the Planck scale it may well be impossible to disentangle black holes from elementary particles. There simply is no fundamental difference. If black holes show any resemblance with ordinary particles it should be possible to describe them as pure states,.....The spectrum of black holes states is discrete. Baryon number conservation, like all other additive quantum numbers to which no local gauge field is coupled, must be violated. [29]

With repeated attempts by various authors, this speculation became more quantified and robust:

> We suggest that the behavior of these extreme dilaton black holes….can reasonably be interpreted as the holes doing their best to behave like normal elementary particles. [30]

A heroic attempt to solve the final state mystery together with the information puzzle was embodied in the concept of final states capable of storing infinite amounts of information—"cornucopions"; it concluded with an attempt to find a dynamic solution for the collapse of a black hole using the string-theoretic equations including a dilaton field:

> We continue to search for a sensible ansatz that will enable us to demonstrate explicitly the existence of a smooth collapsing solution…[31]

What the variable-pole scenario seems to suggest (using a much more naive theoretical structure than the above papers) is that *all* particles may be varying forms of stabilized black holes, and that such objects have appreciable *intrinsic* entropy in addition to any kinematical thermal entropy they might possess. This would put a whole new light on the process of evaporation of large black holes, which might then appear no different in principle from the correlated decays of elementary particles; there would be no need for infinite reservoirs of information. The entropy (and information) could reside just as suggested in [22]—in the deviations of the black hole "atmosphere" from complete thermality—or, more likely, in the structure of the shielding horizons. The large-hole evaporation calculations would have to include a possible flow of correlated information via the intrinsic entropy of the "daughter holes". Without further speculations, we have already seen a big hint that this can happen, namely, in the calculations of Appendix B that showed the new intrinsic entropy alone can satisfy the second law of thermodynamics, for both large and small hole evaporation. Further speculation in Section IV shows how that flow of information might be implemented. Finally, with respect to the entropy behavior shown in Fig. 9a, it appears that this behavior has been expected for some time. QFT leads one to expect a logarithmic dependence of the entropy of its fields on mass, yielding at some large mass to a linear dependence (stringiness), which yields in turn at even larger masses to a quadratic dependence (black holes/strings) [32]. The result of Fig. 9a shows that all of these behaviors follow directly from a single simple model having intrinsic particle entropy.

As of yet, the variable-pole model has no mechanism to induce quantization of the state as it proceeds to arbitrarily low mass, nor to explain the details of the large-$G$ horizon membrane or of a leaky horizon. It suggests, however, that a consideration of unconventionally strong $G$ behavior exceeding that of GR might automatically provide a sensible environment and "gentle landing" for the evolution of an evaporating black hole into any scalar fundamental particle, consistent with QM everywhere and GR on the exterior of the hole. From there, it is only a short flight of fancy to include all elementary particles as stabilized black holes.

More information about this "sensible environment" is contained in Table 1, wherein either the numerical solution or asymptotic form of the solution is used to calculate some physical characteristics of the evaporation.

| Mass (GeV) | $M/M_P$ | $\Delta t$ to Mass (s) | $T_{horizon}$ (GeV) | $G/G_N$ | Entropy (Bits) | $M_{P_v} = M + \delta$ | $R/\lambdabar_c$ |
|---|---|---|---|---|---|---|---|
| $3.66 \cdot 10^{19}$ | $3.0 \cdot 10^0$ | 0 | $1.3 \cdot 10^{17}$ | $1.33 \cdot 10^0$ | 1730 | $(3.66-2.60) \cdot 10^{19}$ | 24.1 |
| $0.61 \cdot 10^{19}$ | $0.5 \cdot 10^0$ | $2.8 \cdot 10^{-40}$ | $2.1 \cdot 10^{17}$ | $5.3 \cdot 10^0$ | 1380 | $(0.61-.075) \cdot 10^{19}$ | 2.61 |
| $0.02 \cdot 10^{19}$ | $1.6 \cdot 10^{-2}$ | $1.0 \cdot 10^{-39}$ | $1.5 \cdot 10^{16}$ | $3.7 \cdot 10^3$ | 1270 | $(0.02+.01) \cdot 10^{19}$ | 1.94 |
| $3.3 \cdot 10^{16}$ | $2.7 \cdot 10^{-3}$ | $5.0 \cdot 10^{-39}$ | $1.3 \cdot 10^{15}$ | $1.4 \cdot 10^5$ | 1200 | $(3.3+.003) \cdot 10^{16}$ | 2.000 |
| $10^3$ | $8.2 \cdot 10^{-17}$ | $\sim 1 \cdot 10^{-25}$ | $4.0 \cdot 10^1$ | $1.5 \cdot 10^{32}$ | 80 | $(1+3 \cdot 10^{-17}) \cdot 10^3$ | 2.000 |
| 1 | $8.2 \cdot 10^{-20}$ | $\sim 1 \cdot 10^{-22}$ | $4.0 \cdot 10^{-2}$ | $1.5 \cdot 10^{38}$ | < 0 | $(1+3 \cdot 10^{-20}) \cdot 10^0$ | 2.000 |
| $10^{-1}$ | $8.2 \cdot 10^{-21}$ | $\sim 1 \cdot 10^{-21}$ | $4.0 \cdot 10^{-3}$ | $1.5 \cdot 10^{40}$ | < 0 | $(1+3 \cdot 10^{-21}) \cdot 10^{-1}$ | 2.000 |
| $0.5 \cdot 10^{-3}$ | $4.1 \cdot 10^{-23}$ | $\sim 2 \cdot 10^{-19}$ | $2.0 \cdot 10^{-5}$ | $6.0 \cdot 10^{44}$ | < 0 | $(0.5+7 \cdot 10^{-24}) \cdot 10^{-3}$ | 2.000 |
| $10^{-9}$ | $8.2 \cdot 10^{-29}$ | $\sim 1 \cdot 10^{-13}$ | $4.0 \cdot 10^{-11}$ | $1.5 \cdot 10^{56}$ | < 0 | $(1+3 \cdot 10^{-29}) \cdot 10^{-9}$ | 2.000 |

*Table 1: Characteristic values of variables in the evaporation of a variable-pole black hole, beginning at $M_0 = 3 M_P$ at t=0. The asymptotic region of the solution is shaded. The amount by which the pole position exceeds the mass M is denoted by $\delta$; note that it changes sign. Where the entropy is < 0, the region is unphysical for scalar particles [see Appendix B]*

While many of the numbers are of unfamiliar scale, they seem to make sense when applied to familiar situations. For instance, the times $\Delta t$ to evolve from a mass $M_0 = 3M_P$ down to familiar particle masses $M$ turn out to be characteristic times for the decay of those particles (if unstable and not constrained by selection rules). The temperatures of the remnant horizons are about 4% of the remnant mass, suggesting a scale of possible binding energy to stabilize that state.

In Table 1, $G/G_N$ has horrific exponents by the time familiar elementary particles are reached, but they are just what is necessary to make gravity comparable to other interactions. The semi-quantitative treatment has shown that a dynamic model of gravity gives it the possibility of different strengths at different mass scales. For instance, an electron internal state (or intra-horizon state) would be characterized by a coupling $G_e \approx 6.0 \cdot 10^{44} G_N$ instead of $G_N$, ignoring for the moment that the electron is not a scalar. This modifies the usual statement of relative strengths of interactions, as in the usual comparison of Coulomb and gravitational forces:

$$F_{gravity}/F_{Coulomb} \approx G_N M_e^2/(\alpha \hbar c) = 2.4 \cdot 10^{-43}$$

(here $\alpha$ is now the fine-structure constant). The variable G-pole model asymptotically has $G_e / G_N \rightarrow (M_P/M_e)^2$, and consequently, if particles could interact with their internal scale of gravity:

$$F_{gravity}/F_{Coulomb} \approx G_e M_e^2/(\alpha \hbar c) = G_N M_P^2/(\alpha \hbar c) \approx 137.$$

Similarly, comparisons of gravity relative to other forces also show this sort of equality. It's the (constant) Planck mass, rather than the mass of some characteristic gravitating particle that is pertinent, if the forces between *unshielded* particles are being compared. Because of the proposed nature of particles themselves, sampling the true gravity forces in nature might seem precluded. However, in Section IV of this paper there are suggestions for experimental tests which might sample and/or falsify such real gravitational strengths, should further speculations advanced there be true.

By far the most dramatic of comparisons in Table 1 is the column where the value of $M_{Pv} = \sqrt{\hbar c / G}$ is calculated for the variable $G$ as a function of $M$. In the asymptotic region (where real particles are) one can again apply the *G*-scaling rule and find:

$$M_{Pv} = M \qquad (!)$$

**If black holes become particles by the mechanism herein proposed, then the Planck scale pertinent to them is their *own mass* at which (for yet unknown reasons) they quantize and stabilize.**

This discussion concludes with the proffered answers to the objections raised at the end of Section I:

1) Gravity is seen everywhere in the Universe because all particles, being bound black hole states of intense gravity, leak the weak asymptotic gravity (GR) into their environs; unstable small black holes and the explosions of the sort sought by experimenters are not necessary for the scenario.

2) Many suggest that gravity is an unshieldable force because there is no matter with repulsive gravity. First note that traditional weak gravity is not shielded in this model, and only a finite strength of strong gravity need be shielded. It is known that about 70% of the universe is made up of a "dark energy" which is very dilute and seems equivalent to a cosmological constant—effectively a repulsive gravity. Is it really certain that the ~4% ordinary matter and the ~25% dark matter could not be shrouded by locally intense concentrations of this unknown quantity? With coarse-graining this might not be noticeable in cosmology, though in fine-grained experiments this mechanism might be falsifiable.

3) As for string theory successes, this paper certainly does not suggest they are wrong or fortuitous; it suggests only that the key elements are the number of degrees of freedom of that theory and the equations, not the interpretation of those degrees of freedom as arising from unseen dimensions, nor the equations as describing real strings. As mentioned, the real shielding mechanism of the strong interactions of QCD arose from the color degrees of freedom and the non-Abelian structure of the theory.  It is possible that the freedom provided by the extra dimensions of string theory map one-to-one onto new aspects of a quantum theory of hypergravity with shielding horizons.  Though not so specifically, some string theorists themselves have pointed out that the detailed concept of strings might be only an artifice to get the right mathematics.  They suggest that the exact duality between some sectors of string theory (ADS space) and ordinary QFT is a two-edged sword which can be taken to be confirming of the reality of strings, or to be indicative of no real need for them to derive the QFT results with which they agree! [32]. The intriguing result mentioned earlier that the entropy of Fig. 9a mirrors QFT, string and black hole theory predictions for entropy (in the mass region appropriate to each) suggests an underlying compatibility among all of these ideas.

4)  Finally, as for the future CERN experiments, they may very well sample a region of $T$ where semi-stable black holes of $M \approx T$ will be produced, but, in the scenario presented in this paper, that is what high-energy physics has been doing all along at other energy scales.  It has been argued that the thermalized evaporation of a TeV black hole would look very different than the fragmentation of a heavy virtual state[33]; this is not at all convincing for a variable-pole scenario.  There, particles/holes decay by passing entropy and information on to the daughter particles/holes, which suggests that strong correlations will exist in the decay products.  From the hypergravity viewpoint, this has always been true in high-energy physics, where experiments repeatedly sampled the appropriate variable Planck "endpoint" for that particular mass region. The scenario offered by this paper, *unlike* that of Giddings and Thomas [2], suggests that the end to conventional high-energy physics is comfortably far away. It agrees with the models of [2] that there is a least size that can be sampled at accelerators.

**Section IV:  Ranked Speculations**

The first three sections of this paper were an attempt to find a quantitative model of small black holes using a crude approximation for conjectured hypergravity.  The result was an internally consistent model which suggests that all elementary particles could be bound states of such a force.[6]  While of course I do not believe such a simple toy model can be representative of the true complexity of quantum gravity, it does suggest that there might be analogous stabilities and feedback within a correct model with an effectively-similar scale dependence of the coupling. Strictly speaking, this should be the end of this paper, and most authors suggesting that particles are black holes eschew any further speculation—enough is enough! For my part, I find it irresistible to ponder what the collateral effects might be in a world in which all fundamental particles are stabilized black holes with the properties the model has described. The following speculations, ranked in order of increasing incredulity, begin conservatively but then extrapolate well beyond the scenario analytically supported by the toy model.

---

[6] From the usual viewpoint of the worker in curved spacetime physics, the very concept of particles is an anathema [35].  The viewpoint espoused here is certainly particle-oriented, not surprising given the occupation of the author.  I assume that the curved-spacetime field theorists, with their ability to rationalize their view with our everyday experience with particles, will continue to do so if a sound basis is found for associating stabilized black holes with elementary particles.

1. Entropy flow and information

Ingenious (and various) mechanisms have been posed to solve the problem of how the entropy gets out of evaporating black holes and thus how loss of information can be avoided [34]. For the variable-pole version of evaporation, it seems it never becomes a problem: black holes evaporate into particles which are stabilized black holes themselves, each with varying but appreciable intrinsic entropy.[7] Using a simplified model to investigate the flow of entropy in black hole evaporation (Appendix B), I found the aforementioned result that the second law of thermodynamics *is satisfied for the intrinsic entropy of black holes alone* over a huge span of masses of the hole. Fig. 12 (App.B) first shows this is true at each step of the evaporation process, with varying efficacy. Furthermore, the flow of total intrinsic entropy is not constant, but starts very rapidly, increasing just as the initial black hole is shedding most of its entropy and then saturating quickly in the sub-Planckian region (Fig.13, App.B). This seems to correspond to the expectations [36], imposed by information flow, for the transfer of entropy necessary to maintain the system in a pure state. At the end of the evaporation, the preponderance of the information has been transferred, and the last few steps near the lowest-lying state proceed just like the usual description of decays of an elementary particle, finishing with a completely coherent non-thermal final state. See App. B and the ensuing discussion there concerning the information flow in very massive black holes. It justifies this information transfer scenario for a scalar world, and I speculate that it continues to hold in general when the model is extended to particles of all spins.

2. The role of GR

Why talk about GR if we are suggesting it needs to be "replaced"? It still governs everything outside horizons, but in my view has a logical extension even there. GR is usually characterized [37] by saying that a given arbitrary initial condition of mass (or stress-energy) tells space-time how to curve and the curvature tells the masses how to move. From my perspective, the masses are likely a consequence of curvature, being made of particles which are local stabilized extrema of the energy of curvature. Then it is natural to think that in some very simple special configurations the mass/energy terms in GR might be replaced by terms also dependent on space-time curvature, leading to the existence of solutions to these modified Einstein equations that would account for both the creation *and* propagation of the masses.  This is hardly a new idea, having been advanced in similar form in 1957 by Misner and Wheeler [46].  The toy model seems very harmonious with a description by geons and GR, effective, of course, only external to the horizons of the black hole/particles.

---

[7] I realize that the toy model is applicable to only zero-spin particles; the extrapolations here are meant to illustrate the principles, not the global picture. If tiny Schwarzschild black holes are spin zero particles with entropy, would anyone think that tiny Kerr-Newman/Reissner-Nordström configurations couldn't be? But the Kerr generalization for entropy would have to have a different form for the M=0 spin 1 or 2 objects, a topic to be considered in further investigations.

3. Black hole interiors

The assumptions of the model suggest that while classical GR will still be valid away from the horizons of small holes, it may have no applicability to their interiors—everything would be "up for grabs", including the metric within. I've proposed a leaky pseudohorizon through which hypergravity transmogrifies into GR, but it is plausible that other aspects of hypergravity "leak out" and generate the other forces and charges of elementary particle theory.  In the simplest conjecture, there is one grand unified force and it is hypergravity.  The toy model brashly predicts that grand unification is indeed at an appropriate Planck temperature, but that it occurs within the interiors of particles at all mass scales.

Let's postulate a "pseudo-singularity" (henceforth termed psingularity) within the horizon—a normal singularity but with the fluctuations expected from quantum gravity. If we wish to keep things as simple as possible, we may conjecture that the interior has a very simple geometry, namely, there is only one coordinate ($\sigma$) within the horizon gauging the distance from the horizon to the psingularity, and every position on the horizon is equivalent to the rest.  This gives us a strange picture of a universe that presumably began at a psingularity containing a host of elementary particles, each of which contains a psingularity—an ugly situation at best.  But an immensely simpler universe results if all of these are the *same* psingularity; everything in the universe is intimately connected through the interiors of the horizons of the particles.  Fig. 10a attempts to illustrate this conjectured metric.  This discussion has relied mainly on intuition, but the picture presented seems to fit well, and extend, the information-conserving model of Horowitz and Maldacena [34].  In that calculation, specification of the quantum state at the singularity was purported to allow teleportation of information across the horizon.  In addition, that model suggested a relation to a possible wormhole nature of the singularity; the extension I suggest is that the wormhole (p)singularities all connect to one another, providing a physical interpretation of the real meaning of teleportation of particle  properties.  What better way to specify the wavefunction at the singularity—those authors even suggest that the boundary condition at the singularity is the wave function of the universe, which would follow naturally for the hypergravity "universal" psingularity.

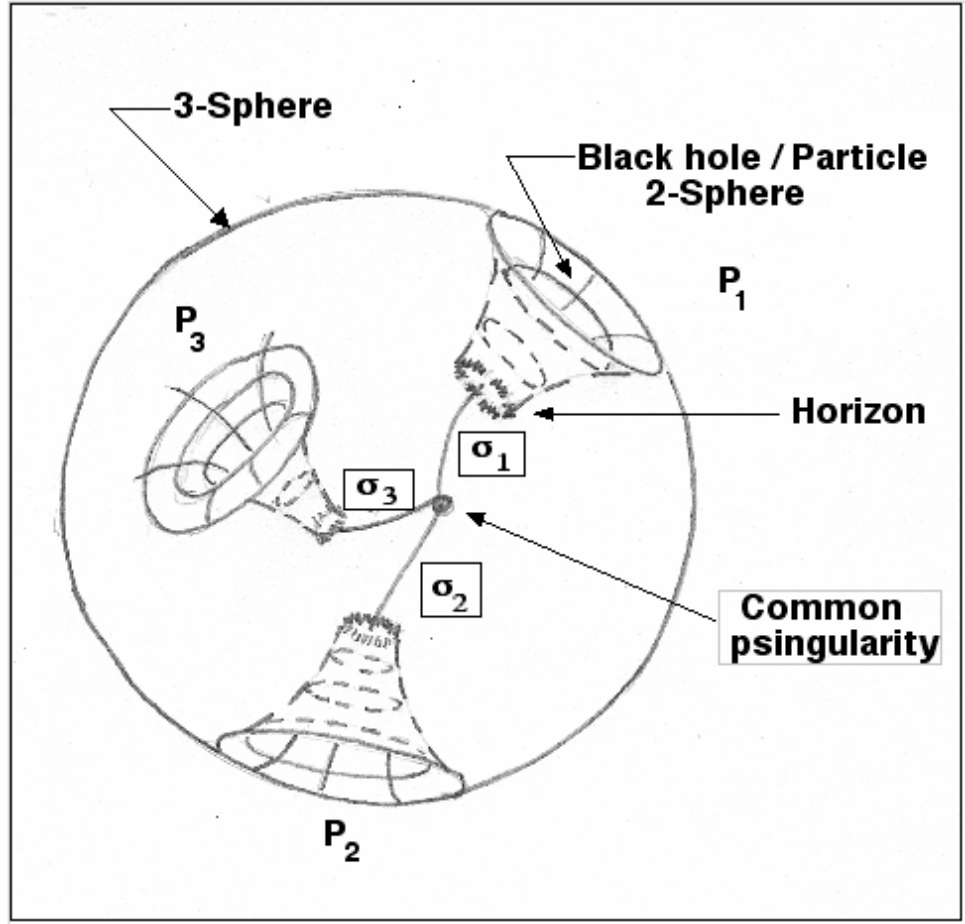

*Fig. 10a: Diagrammatic characterization of a world connected through the internal metric of black hole/particles. The world external to the horizons is represented by the three-sphere, while the particles are represented by their embedding diagrams. Here, $P_1$, $P_2$ and $P_3$ are three coherent particles connected to a common origin by the single dimension $\sigma$ within their horizons.*

What then emerges is a universe filled with stabilized small black holes, which can exchange at least some information via horizon-to-psingularity transfers. One begins to see where non-local effects can come from, except this is a misnomer, because these effects can be truly local in the internal metric. Now, by adding just a bit of complexity, namely multiple steps in that single internal coordinate $\sigma$ to get back to the common psingularity, we can hope to generate most of the features of the non-locality of entangled particles. Fig. 10b, below, attempts to illustrate how two completely entangled particles $P_1$ and $P_2$ might appear, with $\Delta\sigma = 0$ (i.e., in contact or with no intervening vertices), while another particle $P_3$ has effectively disentangled itself by many interactions with other particles since the Big Bang. One might invoke a coupling constant at each vertex in $\sigma$ which would dilute the ability to transfer information, or even interpret the steps as another physical variable (see ranked speculation number 5).

In any case, the peculiarities of the quantum mechanics of highly entangled particle pairs seem much less mysterious in this picture, once one swallows the very large pill of the odd metric, essentially a one-dimensional universal wormhole.

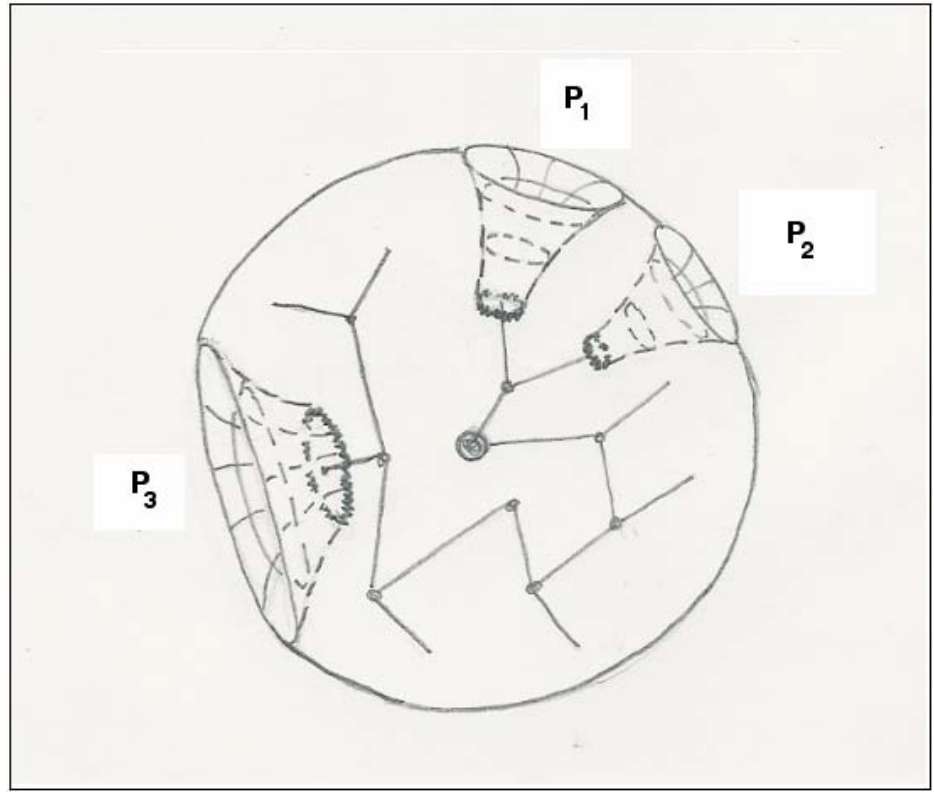

*Fig. 10b: A characterization of possible past histories of the internal metric $\sigma$. Particles $P_1$ and $P_2$ have a recent common past history and are completely entangled, while particle $P_3$ has had other interactions and has at least partially decohered. (The particles in the past history of $P_3$ are not shown explicitly, but are implied by open-ended lines).*

These last few speculations have to inspire one reaction: where is the ability to *falsify* any of the predictions of the model or of the subsequent flights of fancy? Of course I have predicted that there will be *no extra dimensions* and *no anomalous behavior in black hole production* at CERN, but those are absences of effects, not their presence! I've made sharp predictions of the intrinsic entropy of particles and derived the particle temperatures, but see no way to do those experiments. It seems there is only one possibility at present for a real physical effect outside the horizon boundary. Completely entangled particles (σ = 0) might not be shielded from one another's strong gravity, resulting in a distortion of expected forces or dynamics. In the absence of a real theory we cannot make a sharp prediction, but such effects have never been looked for (to my knowledge). An appropriate experimental situation might be one of the following:

(1) *(Likely much too difficult)* Annihilate e+ with e- at rest to produce two entangled photons, analogous to the process illustrated by Fig. 10c. Then try to measure their gravitational redshift as they climb out of the hypergravity potential well: it should be anomalous, probably in the direction of being too large. The generic experiment, to measure the gravitational attraction of entangled fundamental particles, seems nigh impossible to this experimentalist; some other analogous situation might be more accessible.

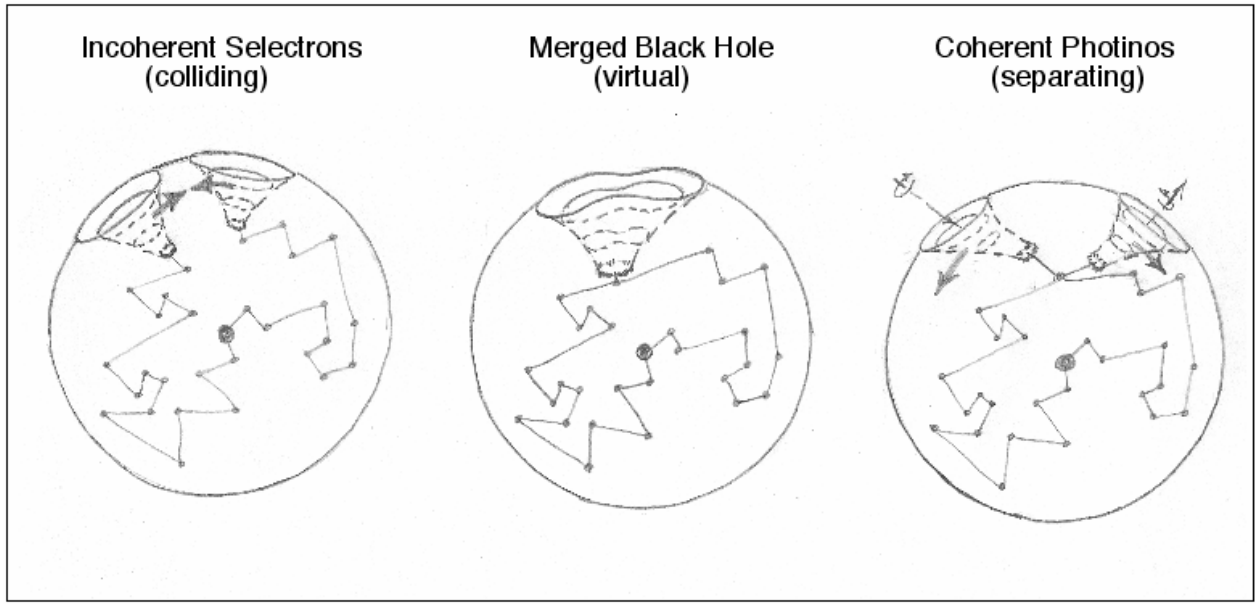

*Fig. 10c: Characterization of a process in which the initial effectively-unentangled state becomes one in which the final state particles might experience an anomalous gravitational force.*

(2) *(Doable but ambiguous)* Recently, a proposal to measure the $G_N$ between entangled particles directly at CERN has been suggested [38]. The scheme uses the decay from an NLSP (next-to-least-mass supersymmetric particle) to a gravitino. Unfortunately, the experiment requires: a) that Nature put the gravitino lowest in mass (stable) among the SUSY particles, and b) that we believe we can correctly calculate a gravitationally-induced decay. As before, we would expect to get something substantially larger than $G_N$.

(3) *(Difficult and ambiguous but underway!)* In a search for possible experiments to support or falsify this model, we happened on an experiment in the planning stage aiming to measure possible breakdowns of Newtonian gravitation at ranges of order 1μ or less. This proposal [39] uses changes in the deBroglie phase between two entangled BEC clouds to sample the changes in gravitational potential (from sample source masses of various sizes and distances). The presence of the Casimir force requires a major background subtraction, which the proposal finds to be feasible. This ingenious proposal suggests to me that such techniques could be used to sample the forces between the clouds themselves, using a third cloud. The problem for our model is that the Rubidium atoms of the clouds are not fundamental, and consist of many levels of particles within particles. Presumably (but not certainly) the constituents have already shielded their anomalous gravity. In fact, if they had not, previous experiments with BEC clouds measuring Casimir forces [40] would probably not have worked. But if my mechanism underlying general entanglement is correct, perhaps even BEC clouds (and other large-scale entanglements) have found a way to communicate through singularities; experimental results would be modified.

4. Cosmologies

One very well known "experiment" presumably sampled the processes described in the first putative experiment above. The Big Bang (some sort of collection of spontaneous curvature fluctuations making many evaporating black holes?) would have produced all known initial particles in just such a sequence as described earlier (I have suppressed the illustration 10d!). The gravity forces may have been quite strongly attractive until the particles exceeded their mean free paths for decoherence. Gravity may have been stronger for some time before it assumed its GR limit; whether any traces of that retardation are present in today's experimental cosmology is unknown. This may be but yet another entry among the non-Newtonian gravity theories trying to account for the unexpected apparent acceleration of the scale factor a(t) of the cosmos. At least it has the correct sign, and could be consistent with $\ddot{a} \approx 0$ at the present epoch. It would, however, predict that a(t) would have no real inflection point; the expansion would simply proceed faster, relative to a model with no dark energy.

Along these same lines of speculation, the manner in which particle theory is used to calculate the cosmological constant (with disastrous results) indicates that there may be something very wrong with the way we calculate vacuum energies. The loop calculation involving gravitons, which is generally considered the culprit causing the introduction of a high power of the Planck mass, might be very different if every particle, virtual or otherwise, is cloaked with a repulsive or shielding component of gravity. We can't even be sure what a graviton might be, in such a theory.

If hypergravity and speculation (3.) are correct, they imply a new viewpoint of the connectedness of the Universe, this seems to supply grist for old and new mills alike: Is there a new way to invoke Mach's principle for the interpretation of inertia that could be more quantitative than in the past? Is the horizon as an area[8] for information storage (for communication through the psingularity) the reason why black holes seem to have a holographic nature [34], and why there seem to be mathematical connections in particle theory between the very small and very large distance scales? (Recall that the attempt to make a small mass black hole resulted in an object which expanded like its Compton wavelength; there *were* no smaller distance scales than the Planck scale.) And, would the conjecture that externally causally-disconnected but internally coherent particles near the start of the Big Bang might have exchanged "non-local" information change our picture of why the Universe is isotropic? The hypergravity picture of the cosmos can raise all these questions but of course at this stage can answer none of them.

5. Time

I began by hoping to elucidate why gravity seems weak, without invoking new dimensions, and seemed to have added a new internal dimension $\sigma$ which proceeds in steps. As a final speculation, I associate this dimension with time.[9] Then we may interpret the world as a series of black hole/particle configurations, separated by the $\sigma$-steps necessary to make an elementary change of configuration. The propagation of a single isolated particle would then be interpreted as the particle communicating with identical virtual particles (the standard concept of multi-path amplitudes), constrained by what previously had determined its momentum, and in one $\sigma$-step developing an amplitude for the next position of the particle. For the two-slit experiment, the unconstrained photon propagates everywhere and interferes, while the application of constraints to the photon (experiments) causes decoherence of all the amplitudes because of the "non-local" information flow. The possibilities for quantization of space and time, and for an upper limit to velocity are obscure but intriguing.

From such a speculation, it is manifest where the connection between the direction of time and the thermodynamic arrow could really arise. Two particles which have decohered, many σ-steps apart, are those which have interacted many times (increasing the net entropies) and have long disjoint time histories. If they are forced to "time-reverse" their histories by new interactions (and further increase of entropy) with other particles, they don't go backwards in time, but even further forward, and with more entropy produced.

---

[8] There is no volume of a black hole in this picture.

[9] In GR, the time and space coordinates interchange roles inside the horizon, while in the spacetime external to horizons, the metric demands that time is treated differently than the other dimensions; perhaps these effects can be considered a consequence of making relativity theory "hook on" to a radically different spatial metric that exists within the horizons of particles.

Puzzles:

Speculative papers such as this often promise resolution to every problem in physics; I hasten to say that the only *real* result of this paper is that a particular form of hypothesized *G* stabilizes the endpoint of the classic Hawking solution for evaporation and predicts intrinsic entropy.  Even the rank(ed) speculations leave many problems unsolved, and for humility's sake, I list some favorites.

Antiparticles:   A satisfactory world-view should give a real insight into the nature of antiparticles.  It is insufficient to just expect them to appear as an extra solution to an eventual complete mathematical formulation of gravity, as they do in the Dirac equation for QED.  The fact that they formally appear in QFT as particles moving backwards in time suggests that they correspond to some reversed dynamic process on the σ-parameter in the above speculations.  This is hardly helpful, except to suggest that antiparticles may be stabilized black holes coming from the future. Such particles would provide a real mechanism for advanced action and presumably make almost everyone unhappy except for one well-known philosopher [41].

Quantization:   One wants a correctly predictive reason why the evaporation process stops (almost) at particular semi-stable states corresponding to elementary particles.  Evaporation in the model would slow down because $\alpha$ falls back through the steps (in Fig. 1) as the temperature of the hole decreases (a complication not included in our model, which will simply stretch the time scale). Evaporation stops when $\alpha = 0$, i.e. when there is no set of daughter particles that honors all the known conservation laws; but evaporation into multiple infrared photons seems to let these black holes evaporate forever.  The positive specific heat allows this kind of hole to come into equilibrium with the microwave background at a neutrino-like mass of 6 meV, and at higher masses when the universe was hotter. While interesting, this is not true quantization.

Big horizons:  If particles are really pseudo-horizon structures as large as the Compton wavelength of the particle, then why do interactions give pointlike sizes: e.g., why is the electron size $<10^{-5}$ f in interactions while $\lambdabar_{c_e} \sim 400$f?   If it has something to do with the interaction of two pseudo-horizons, what changes the scale so much?   And if a giant black hole has a GR horizon that is non-material, how does such a horizon become the substantial one as evaporation proceeds, or as one falls in (with no effect) and then approaches the singularity?

These questions, and likely many others, would need resolution in a reputable theory of hypergravity.

Conclusions:

We find that many puzzles of modern particle/gravity physics have a natural qualitative explanation in a radically different world-view of the nature of elementary particles/black holes. While only a scenario, it is surprising that many of these explanations also follow quantitatively from a simple toy model that invokes only one new feature: a proclivity of gravity to become very strong at small black hole masses, but also allows it to be dynamically self-quenched.

Rather than competing with other models and theories, we can look for connections between our results and the "new paradigm" of black hole complementarity, holography and the yet-unknown microscopic degrees of freedom of quantum gravity [42]. The flow of entropy of entanglement out of an evaporating black hole seen in our model, plus the possible non-local nature of the model, certainly is parallel in spirit if not in mathematical development with these other ideas. The model suggests that Susskind's unknown microscopic degrees of freedom are the incredibly complex horizons which might act as selective filters for our already-discovered force fields. The accompanying speculations also provide possible connections among quantum non-locality, holography and the short/long-range dualities[10].

We would hope that this world-view finds further validation in a yet-to-be-discovered sophisticated theory incorporating these viewpoints.

---

[10] After deriving the results of this paper, we discovered literature suggesting some of these effects in papers we had missed, using more sophisticated treatments. This is noted just to reassure the reader that the model was not generated to produce these results—they were all a surprise. These papers are referenced in the text.

# APPENDIX A:  Asymptotic behavior of the evaporation variables.

We know from the form of the solution that *M* can only decrease with time, and that it must tend toward zero.  From **(11)**, we see that the $M_0/M$ terms will eventually dominate, and that after some time $t_a$, the value of *G* will be arbitrarily close to $G_N(M_P/M)^2$.  Substitution of this value of *G* into **(4)** gives : $\dfrac{dM}{dt} = -\dfrac{\beta M^2}{(\hbar c)^2}$, which immediately leads to the analytical solution $\Delta t \equiv t - t_a = \dfrac{(\hbar c)^2}{\beta}\left(\dfrac{1}{M} - \dfrac{1}{M_a}\right)$.  After some manipulation, this takes the form: $M = \dfrac{M_a M_P^4}{M_P^4 + \alpha M_a \Delta t}$, where $\alpha$ was defined after **(4).**

For $\Delta t$ sufficiently large, this gives the simple form: $M = \dfrac{M_P^4}{\alpha \Delta t}$.  We find that the region of validity of this approximation is for $\Delta t > \sim 1.4 \cdot 10^{-40}$s, or, for the solutions shown that start at $M_0 = 3M_P$, for $t > \sim 2 \cdot 10^{-39}$s (that is, about twice the maximum time shown on the plots).  By $t > 2 \cdot 10^{-38}$s, the precision of the approximation is about 0.1% — that's at a mass of about $10^{16}$ GeV, quite adequate for our interests concerning elementary particles.

Applying this approximation in turn to the defining relations for *G*, *T* and *R*, we summarize the results as scaled variables:

$$\frac{M}{M_P} = \frac{M_P^3}{\alpha t} \qquad \frac{G}{G_N} = \frac{(\alpha t)^2}{M_P^6}$$

$$\frac{T}{T_P} = \frac{M_P^3}{\alpha t} \qquad \frac{R}{R_P} = \frac{\alpha t}{M_P^3}$$

From these, we arrive at the useful interrelations used in the text, valid for small M:

$$G/G_N = (M_P/M)^2 = (R/R_P)^2 = (T_P/T)^2.$$

If we also note that for large M, the last three relations are:

$$(M/M_P) = (R/R_P) = (T_P/T),$$

then the inherent duality properties of the variables under the exchange $M \leftrightarrow 1/M$ are manifest.

There remain two curiosities to clear up — we hypothesized a G which was supposed to increase monotonically with T, and chose a formulation guaranteeing this equation **(1)**. With the change to **(9)**, we lost this assurance, and asymptotically found $T \sim 1/t$ while $G \sim t^2$, clearly contradicting the original idea. We can live with the idea that the model had better intuition than we did, but there is still a puzzle. In Eq. **9**, applying these asymptotic behaviors, it appears that:

$$G \to \frac{T_P - at}{T_P - b}, \quad \text{where } a \text{ and } b \text{ are constants.}$$

Then G appears to go to zero or even negative as $t \to \infty$. Evaluating the constants shows that in actuality, the limit looks like $G \to 0/0$, and proper evaluation of the indeterminate form shows that $G \to \propto t^2$, so all is well.

But there is another place this formula is curious, in that a cancellation of the pole in G occurs. This happens at $M = M_P/4$ for our particular solution, where the mass of the evaporating hole actually "catches up" with the pole position:

$$\frac{G}{G_0} = \frac{T_P\left(1 - \frac{T_0}{T_P} 4\right)}{T_P\left(1 - \frac{1}{4} 4\right)}$$

But no infinity in G need occur, because $\frac{T_0}{T_P} = \frac{M_P G_N}{M_0 G_0} = \frac{1}{3} \cdot \frac{3}{4} = \frac{1}{4}$ and the numerator also vanishes. A careful look at the proposed $G/G_0$ shows it to be of the form $y = \frac{u(x)}{1 - f(x)/y}$, where $y = G/G_0$, and $x = M/M_0$ ($= 1/12$ at the "pole"). Thus we get $y = u(x) + f(x)$, or at the "pole" $y = f(x)$ and is not singular; $u(x) = 0$ at the pole. Evaluation at the "pole", showing the explicit form of $f(x)$ gives:

$$y_{pole} = \frac{1}{2x_{pole}} \frac{M_P^2}{M_0^2} \frac{G_N}{G_0} \left\{ \left(\frac{1}{x_{pole}} - 1\right) + \sqrt{\left(\frac{1}{x_{pole}} - 1\right)^2 + 4\frac{M_0}{M_P}\left(1 + \frac{M_0}{M_P}\right)} \right\}$$

$$= \frac{1}{2}(12)\frac{1}{9} \cdot \frac{3}{4}\left\{(12 - 1) + \sqrt{(12 - 1)^2 + 4 \cdot 3(1 + 3)}\right\} = 12$$

So, at the pole, the "pole" disappears! Even stranger, for all $t > t_{critical}$ (at which the "pole" is reached), the variable pole $M_{Pv} > M$, but in addition $\delta = M_{Pv} - M = 0.33$ $M^2/M_P$. The pseudo-pole is forever locked to the value of the evaporating mass! This is why the solution for $M(t)$ has such unique and dynamic properties.

# APPENDIX B:  Entropy flow in evaporation/decay.

We wish to calculate the net gain or loss of entropy in a black hole-thermal environment when the black hole evaporates, counting only the intrinsic entropies possessed by the original hole and the emitted particles. If we were to do this classically, we would find a marked *decrease* in the total intrinsic entropy, as exemplified by the following simple case: a mass of $M_o$ completely evaporates into $N$ identical particles at rest, each of mass $M_i$ with $N = M_o/M_i$. Tracing the intrinsic entropy $\mathbf{S}_o = 4\pi G_N M_o^2/\hbar c$, $\mathbf{S}_f = N(4\pi G_N M_o^2/N^2)/\hbar c$, we see that $\Delta\mathbf{S}_{gain}/\Delta\mathbf{S}_{loss} = 1/N$. So for any $N > 1$, the intrinsic entropy decreases, and for the usual case where $M_i \ll M_o$, it decreases enormously.

In the real world, large black holes are not expected to bifurcate or polyfurcate into moderate sized holes, but instead into particles *elementary at the scale of the large black hole temperature*. Fig. 1 showed what this means quantitatively. Fig. 11 shows an interesting duality symmetry we noticed long ago [43] between the mass of holes and the heaviest particle into which they can decay. One sees immediately that the simple example above has no realization unless $N$ is gigantic. Moreover, even though GR and the variable-pole model give identical results for the intrinsic entropy in the large mass region, the *flow* of entropy will be entirely different for the two schemes, because the larger masses reference the sub-Planckian region where the treatment of GR/Hawking objects and variable–poles has to be completely different.

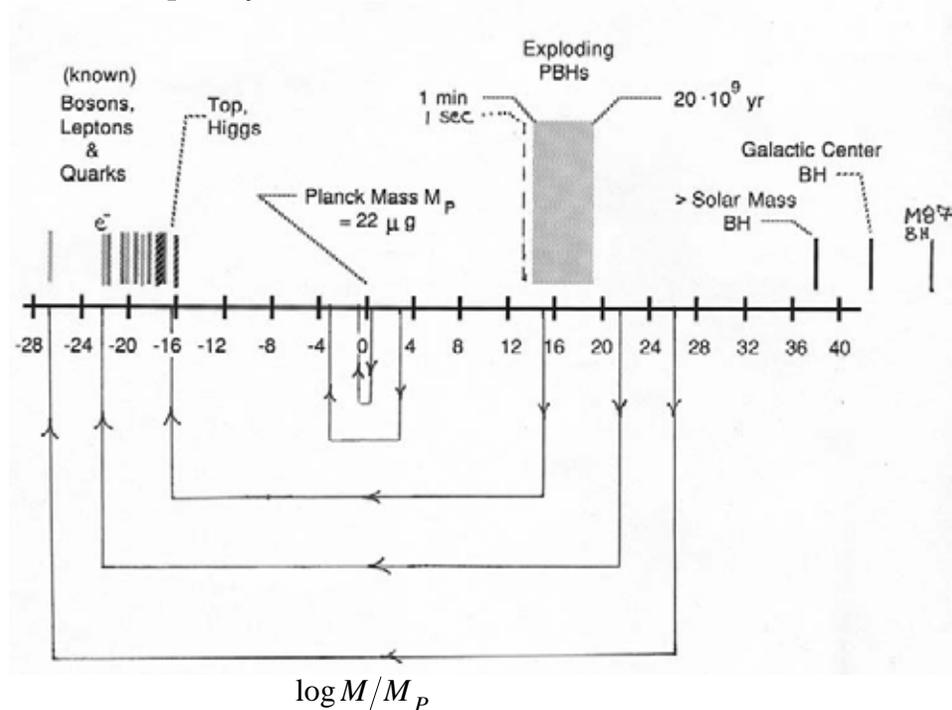

*Fig. 11: The landscape of masses M: the arrows indicate the most massive particle (at left) that can be emitted by a black hole (at right).*

Thus, in the case of Hawking evaporation of a large hole, the only entropy flow is from the intrinsic large-hole entropy into the non-intrinsic entropy of fundamental particles (reflecting the kinematical information about their micro-canonical ensembles). In our new world view, the final state particles have intrinsic entropy too, in addition to kinematical information content. This intrinsic entropy could in principle be entropy of entanglement, i.e., it could be combined with the entropy of the evaporating black hole and other particles to keep the combined state a pure one. (This assumes the original black hole was in a pure state). For this view to be a real possibility, the intrinsic entropy in the final state should *increase* steadily throughout most of the evaporative process, which is just the opposite of what we found in the simple classical case. So, what really happens?

A proper calculation of the evaporation process, whether it starts at super-$M_P$ or sub-$M_P$ masses, is beyond the scope of this paper. It would involve a large Monte Carlo program as described in [44], modified for the variable-pole assumptions—convolving all possible particle distributions (production, fragmentation, decay) with the Hawking radiation equation. We can, however, do an approximate calculation more sophisticated than the trivial, but revealing one, with which this Appendix began.

A key approximation, justified by [45], is that one can approximate the Hawking thermal spectrum by a δ-function at the peak energy of emission, $\sim 5k_B T = 5\kappa k_B/MG_N$, which then becomes the typical energy per emitted particle. Then, instead of a one-step process, we take the more precise view of the evaporation proceeding differentially. We assume the hole of initial mass $M_o$ makes a small step of $-\Delta M$, and that energy $\Delta M c^2$ is divided among $\Delta N$ similar particles $M_i$ at rest, all of which have the typical energy. That gives us $\Delta N = \dfrac{\Delta M c^2}{E_{Typ}} = \dfrac{8\pi}{5M_P^2} M \Delta M \left[ \dfrac{G(M)}{G_N} \right]$, where we have kept $\Delta N$ general by leaving $G$ variable instead of using $G_N$. We designate the two different asymptotic regions of mass above and below $M_P$ by the subscripts "sup" and "sub" respectively. Then, when the mass $M$ is in the super-Planckian region, we have:
$$_{sup}M_i = \frac{\Delta M}{_{sup}\Delta N} = \frac{5M_P^2}{8\pi M}, \text{ because } _{sup}\Delta N \text{ has } G = G_N.$$
Correspondingly, when the mass $M$ is in the sub-Planckian region, we have:
$$_{sub}M_i = \frac{\Delta M}{_{sub}\Delta N} = \frac{5M}{8\pi}, \text{ because } _{sub}\Delta N \text{ has } G = G_N \left(\frac{M_P}{M}\right)^2.$$
(Note that in the far "sub" region, $M <<< M_P$, the multiplicity $_{sub}\Delta N$, for a given $\Delta M$, is larger, reflecting the fact that the hole temperature has become low again and the typical energy is thus much smaller). Now, for each region we calculate a "figure of merit" $R$, which tells us whether we are gaining or losing intrinsic entropy for that $\Delta M$.

$$R = \frac{\Delta S_{\text{gained in final state}}}{\Delta S_{\text{lost from } M_0}}, \text{ where } {}_{\text{sup}}S = S'_L + 4\pi\left(\frac{M}{M_P}\right)^2, \text{ and } {}_{\text{sub}}S = S_L + 8\pi \ln\left(\frac{M}{M_L}\right)^2.$$

$S_L$ is the intrinsic entropy of the lowest scalar particle which has mass $M_L$. $S'_L$ is the appropriate constant of integration for the large-mass entropy; it cancels out in the differential. One must keep close track of whether $\Delta N$ comes via a mother from the sub or sup regions, and whether $M$ is the mother or daughter mass. If one perseveres, the resulting formulas give

$$_{\text{sup}}R = \frac{1}{5}\left\{S_L + 8\pi \ln \frac{5}{8\pi} \frac{M_P^2}{M_L M}\right\}, \text{ if } M \gg M_P;$$

$$_{\text{sub}}R = \frac{1}{5}\left\{S_L + 8\pi \ln \frac{5}{8\pi} \frac{M}{M_L}\right\}, \text{ if } M \ll M_P.$$

Remarkably, these formulas are rather simple and possess the same symmetry as that exhibited in Fig. 11. The only parameters of this formulation are the mass and intrinsic entropy of the lowest mass scalar particle. We take two broadly different values for each and plot the results for the differential flow of intrinsic entropy in Fig. 12 below.

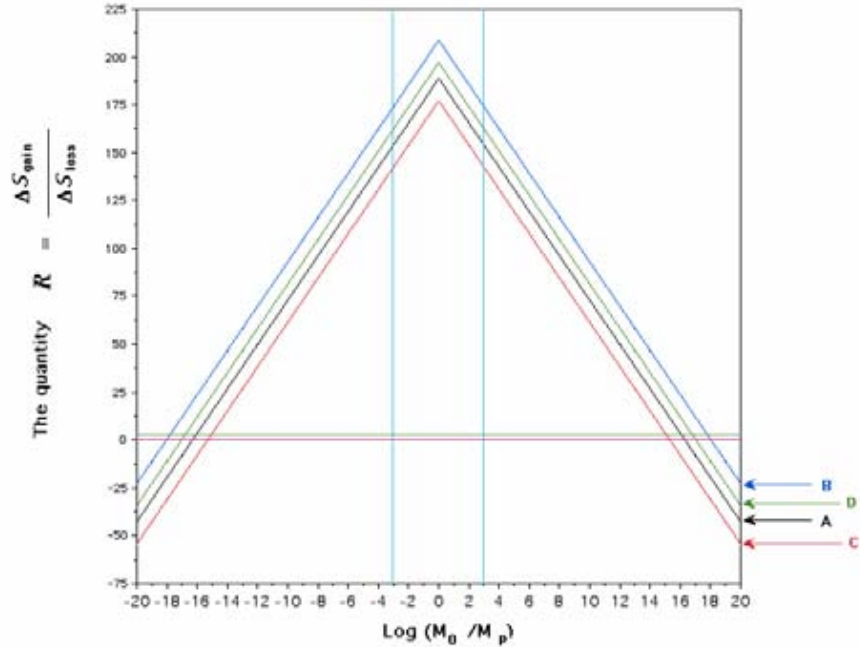

Fig. 12: The regions of increase (R>1), decrease (0<R<1), and unphysical (R<0) for intrinsic entropy in the variable-pole model, with lowest scalar at:
A) 100 GeV, entropy 0; B) 100 GeV, entropy 100;
C) 1 TeV, entropy 0; D) 1 TeV, entropy 100

In Fig. 12, the region between the two vertical lines indicates the transition region where the asymptotic solutions become inaccurate. [A detailed treatment would simply have a line of positive slope curve into one of negative slope (as we saw in Fig.6); the precise result is not necessary for our present argument.] The plot shows that over 32 orders of magnitude, the differential gain in intrinsic entropy from the decay exceeds the loss, and greatly so in most of the regions and for all the variation of parameters. From this result we can expect that the Second Law will be satisfied for the intrinsic entropies in essentially all cases.

This conclusion can be made more explicit and further insights gleaned, just by integrating the sum of the differential changes in intrinsic entropy found above:

$$\mathbf{S}_{total} = \int_{M_0}^{M} \left( \Delta \mathbf{S}_{\text{gained in final state}} - \left| \Delta \mathbf{S}_{\text{lost from } M} \right| \right)$$

We display $\mathbf{S}_{total}$ in Fig. 13 (next page), using a variety of initial masses $M_0$. ($\mathbf{S}_L$ is taken to be zero, but even with $\mathbf{S}_L = 100$, the general nature of the curves in Fig.13 does not change—the saturation values for the total entropy just rise somewhat higher). One sees that the evaporation starts at the right of each curve, with the initial intrinsic entropy marked by a small circle. In every case for mother holes of super-Planckian mass, the total intrinsic entropy rises sharply as the evaporation proceeds along the direction of the arrows. The entropy saturates quickly for all cases except for mother holes of sub-Planckian masses, where the rise and saturation occur more slowly. The integration is terminated (at the left side) when the peak energy of Hawking radiation becomes smaller than $M_L c^2$ and the Halzen approximation breaks down. By this point the remaining mass and entropy of the mother black hole are very small, and the evaporation is best described by ordinary particle decay.

By contrast, if we still use the same approximation, but say that all black holes—including sub-Planckian ones—have the classic intrinsic entropy accorded to large black holes, we get the dotted trajectory for total entropy shown below in Fig. 13. The Second Law is violated quite severely. It is the logarithmic behavior of particle intrinsic entropy that is "saving" the Second Law. Also note that just as the mother hole is rapidly shedding its entropy, proportional to $M^2$, the daughter particles go off with even more than was lost. This timing was not evident in Fig. 12, where the figure of merit peaked much later, near the "phase transition".

But what happens if a huge black hole is evaporating—is not all the information lost before the mass shrinks to ~$10^{16} M_P$? In the strictly scalar model, there are no particles with $M < M_L$, so black holes above $10^{16} M_P$ don't evaporate at all—there are no γ's or ν's! Here we discover the limitations of the scalar model. But even as derived, there is a strict correlation between the lowest mass $M_0 \equiv \eta_{sub} M_P$ for which $R = 1$, and the highest mass $M_0 \equiv \eta_{sup} M_P$ for which $R = 1$. One discovers that

$$\eta_{sub} = \frac{8\pi\varepsilon}{5} e^{(5-S_L)}, \text{ while } \eta_{sup} = \frac{5}{8\pi\varepsilon} e^{-(5-S_L)}, \text{ (where } \varepsilon \equiv \frac{M_L}{M_P}\text{),}$$

implying the duality and symmetry we saw earlier, that:

$$\eta_{sup} = \frac{1}{\eta_{sub}}.$$

Thus even in the scalar model, if $M_L \to 0$ ($\varepsilon \to 0$), the $\eta_{sub}$ and $\eta_{sup}$ endpoints ($R=1$) move to 0 and $\infty$ respectively, and $R \geq 1$ for black holes of any mass. I speculate that this will still be true if the model were generalized to include vector and spinor black holes, eliminating any case which would have information loss in the evaporative process. It is perhaps unwise to pursue the model in so much detail. But overall, the conclusion is simply that there seems to be a great deal of previously uncounted intrinsic entropy in this model.

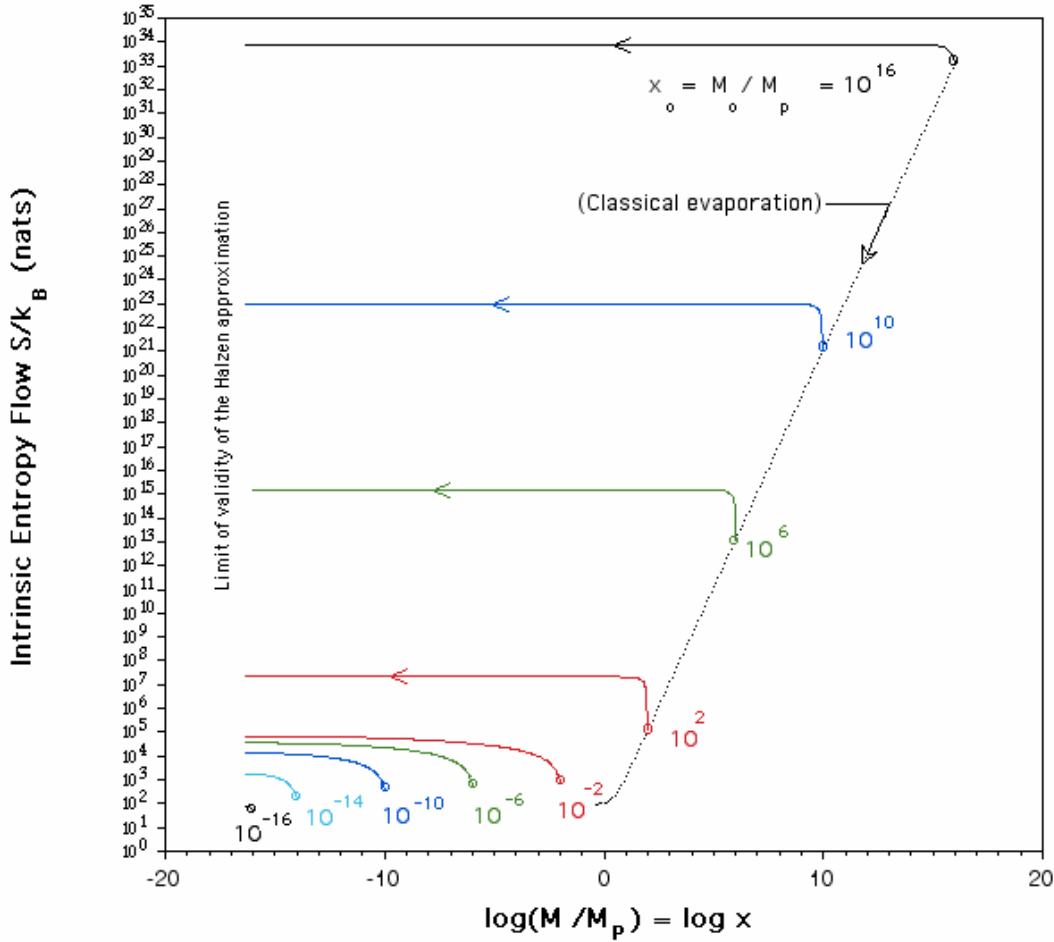

Fig. 13 The change in total intrinsic entropy as a variable-pole hole evaporates: parameter is initial mass; classical solution shown by dotted line.

# APPENDIX C: An alternative construction of the model.

There is something definitely fishy about the form used for *G* in equation **(8)**: it seems to have no parameters to characterize the rapidity or strength with which *G* approaches the "pole". In fact, it does, and such a parameter is embodied in the condition $\frac{G_0}{G_N} = 1 + \frac{M_P}{M_0}$. The time solution will thus be different for every different starting value $M_0^*$, even after the mass decays down to the value of $M_0$ we used! (This can be easily proven by examining equation **(12)** with two different starting points). For initial starts near the Planck mass, e.g. $M_0 = 3\, M_P$ or $M_0^* = 10\, M_P$, the solutions are trivially different and the general conclusions of the text are invariant. But for a start at $M_0^* = 10^{45}\, M_P$, gravity may be too weak to ever cause a transition at Planck-scale masses. This stems from having the strength parameter implicitly defined in the expression for *G*.

Are there other forms of *G* which have these desired properties, while eliminating the ambiguities of the embedded intital conditions? Yes, there appear to be many such formulations, but they seem to have a common condition—all needing a form depending on, or equivalent to, a variable pole in T. Forms with exponentially-increasing G or even simple poles do not exhibit the phase transition and consequent behavior. A choice that appears to be better than the original one chosen in the main text of this paper was suggested by paper reviewer David M. Cheng and modified by me:

$$G(T) = G_N \frac{T_{P_v}}{T_{P_v} - \mu T} \;, \quad \text{where } \mu \text{ can be called a "pole-onset" parameter.}$$

The expression automatically satisfies $G(0) = G_N$. It has no solution ambiguities, because initial conditions appear normally as limits on integrals, not embedded in the definitions. It also has the free parameter $\mu$, which if $\mu < 1$ delays the onset of the phase transition to arbitrarily low masses. If $\mu > 1$, the phase transition happens very early, "freezing" the large black hole at a high mass $\gg M_P$, and effectively stopping evaporation. Cheng framed his version of this definition using no $\mu$ parameter (or $\mu = 1$), and the resulting solutions come out almost exactly duplicating those of this paper, with numerical locations of various features displaced by a maximum amount of order 10%. Finally, this generalized form has the attractive feature of being analytically integrable; equation **(12)** of the text becomes, with *x* defined as $x = M/\mu M_P$,

$$t - t_0 = \mu^3 \frac{M_P^3}{16\alpha} \left\{ \begin{array}{l} \frac{16}{3}(x^3 - x_0^3) + 32(x - x_0) - 8\left(\frac{1}{x} - \frac{1}{x_0}\right) + 8\sqrt{4 + \frac{1}{x^2}}(x^2 - 1) \\ -8\sqrt{4 + \frac{1}{x_0^2}}(x_0^2 - 1) + 20\ln\left(\frac{x}{2}\sqrt{4 + \frac{1}{x^2}} + x\right) - 20\ln\left(\frac{x_0}{2}\sqrt{4 + \frac{1}{x_0^2}} + x_0\right) \end{array} \right\}.$$

The basic scenario for strong gravity remains the same as presented earlier.

## Acknowledgements

I had a great deal of help from Jonathan Michael Hanna. He performed repeated numerical calculations as part of his UCSC Senior thesis [17], supervised by me. He also checked all of the tedious algebra I had done in arriving at the equations to be numerically solved. It is also noteworthy that as we struggled to find an appropriate form for *G*, we simultaneously and independently both arrived at the key concept of the variable Planck mass.

Many thanks to physicists David Cheng, Fred Kuttner and Bruce Rosenblum, who helped with detailed checks of the paper. To say that they helped is not to say that they always, or ever, agreed.

As I was learning my way into some of the arcane topics of this study, I had the opportunity to discuss the general topics with physicists Tom Banks, Francis Halzen, Gerard 'tHooft, Michael Nauenberg, Richard Price, Joel Primack, Matthew Sands, Bruce Schumm, Abraham Seiden, Frank Wilczec and Robert Wald, and I thank them for their help. They bear no responsibility whatsoever for any errors of interpretation I might have committed herein.

I also appreciate the encouragement and support given to me by Lelia & Gwendolyn Coyne, Paul and Irinel Petrescu, Melanie Mayer, and Ruth Valdez, some of whom also gave me the benefit of non-physics critiques of the construction of the text.

The U.S. government and the National Science Foundation will likely be relieved to learn that no grant monies were used to support this work. I do thank the University of California Retirement System for funds used to support me while completing this project, even though it would have been illegal for it not to have done so.